\documentclass[useAMS,usenatbib]{mn2e}

\usepackage{hyperref}
\usepackage[fleqn]{amsmath}
\usepackage{amsfonts}
\usepackage{graphicx}
\usepackage{amssymb}
\title[Extreme value statistics of weak lensing shear peak counts]{Extreme value statistics of weak lensing shear peak counts}
\author[R. Reischke et al.]{R. Reischke$^{1}$\thanks{E-mail:
reischke@stud.uni-heidelberg.de}, M. Maturi$^{1}$, and M. Bartelmann$^{1}$\\
$^{1}$Zentrum f\"ur Astronomie, ITA, Universit\"at Heidelberg, Philosophenweg 12, D-69120, Heidelberg, Germany}
\begin{document}

\date{Accepted 2015 November 12. Received 2015 October 12; in original form 2015 July 08
}

\pagerange{\pageref{firstpage}--\pageref{lastpage}} \pubyear{}
\maketitle

\label{firstpage}

\begin{abstract}
The statistics of peaks in weak gravitational lensing maps is a promising technique to constrain cosmological parameters in present and future surveys. Here we investigate its power when using general extreme value statistics which is very sensitive to the exponential tail of the halo mass function.\\
To this end, we use an analytic method to quantify the number of weak lensing peaks caused by galaxy clusters, large-scale structures and observational noise. Doing so, we further improve the method in the regime of high signal-to-noise ratios dominated by non-linear structures by accounting for the embedding of those counts into the surrounding shear caused by large scale structures.\\
We derive the extreme value and order statistics for both over-densities (positive peaks) and under-densities (negative peaks) and provide an optimized criterion to split a wide field survey into sub-fields in order to sample the distribution of extreme values such that the expected objects causing the largest signals are mostly due to galaxy clusters. We find good agreement of our model predictions with a ray-tracing $N$-body simulation.\\
For a Euclid-like survey, we find tight constraints on $\sigma_8$ and $\Omega_\text{m}$ with relative uncertainties of $\sim 10^{-3}$. In contrast, the equation of state parameter $w_0$ can be constrained only with a $10\%$ level, and $w_\text{a}$ is out of reach even if we include redshift information. 
\end{abstract}


\section{Introduction}\label{sec:1}

The cluster mass function has proven to be a powerful tool to constrain cosmological parameters (e.g. \citealt{2001AIPC..586..303H, 2004ApJ...613...41M, 2004PhRvD..70l3008W} and \citealt{2009ApJ...692.1060V}). However, the mass is not a direct observable and can only be inferred indirectly. Not even gravitational lensing provides a direct measure of mass because it only constrains the gravitational potential, which is a local quantity in contrast to the mass. It is therefore advantageous to replace the mass with a well defined observable quantity to constrain cosmological parameters in an unambiguous way. Additionally, this observable needs to be predictable from theory in a reliable and clear way. There are different ways to tackle this problem. For instance the mass can be replaced with the X-ray temperature of galaxy clusters (\citealt{2009A&A...494..461A, 2015arXiv150403187A}). 

In this paper we replace the mass with the signal-to-noise ($S/N$ henceforth) ratio of weak lensing shear detections (\citealt{maturi2009anaapptonumcouweapeadet, 2011MNRAS.416.2527M}, which is based on the work of \citealt{bardeen1986stapeaGauranfie} and \citealt{2000MNRAS.313..524V}). We then apply general extreme value (GEV) and order statistics to the abundance of $S/N$-ratio peaks in cosmic shear maps to reproduce the distribution of the highest detections (i.e. with the highest $S/N$) correctly. Cosmic shear peaks originate from three sources: i) From noise due to the intrinsic ellipticity distribution and the finite number of sources, ii) from chance alignments of large-scale structures (LSS) and iii) from cluster-sized dark matter halos. The highest peaks in a weak lensing map will be due to clusters. Therefore they will carry similar cosmological information as the mass function. Given the constraining power of weak lensing peak counts (\citealt{2009ApJ...698L..33M}, \citealt{PhysRevD.81.043519} and \citealt{2010MNRAS.402.1049D}), the abundance of weak lensing detections has been predicted and used in 
\citet{maturi2009anaapptonumcouweapeadet}, \citet{2011MNRAS.416.2527M}, \citet{2013MNRAS.430.2896C}, \citet{2014arXiv1409.5601C}, \citet{2014PhRvD..90l3015P}, \citet{2014JCAP...08..063M} and \citet{2014arXiv1410.6955L}. 

Under-densities (voids) produce weak lensing signals as well as over-densities. It is possible to model the abundance of negative detections by a Gaussian random field, as those objects evolve mostly linearly. In contrast the over-densities produce detections which are due to halos and therefore due to highly non-linear objects.

We use the abundance of detections of both positive (over-densities) and negative (voids) to extract their respective GEV and order distributions. This was already done for the mass function in \citet{2011MNRAS.413.2087D}, \citet{2011MNRAS418456W}, \citet{2012MNRAS4201754W} and \citet{2013MNRAS.432914W}.
This approach has the advantage that only the highest weak lensing peaks are taken into account which are prominent features in weak lensing maps. Even though information is in principle lost when considering the highest peaks only, these objects will populate the exponential tail of the mass function and are therefore very sensitive to variations in the cosmological parameters. Furthermore this approach avoids the assumption of a Gaussian likelihood and links the observable directly to its distribution.

We focus on the statistical distribution of the highest positive and negative weak lensing peaks, which trace the most massive objects and the most under-dense regions respectively. This approach has some advantages: First  identifying weak lensing detections, i.e. connected regions that have a weak lensing signal larger than some threshold, allows to avoid the poorly constrained mass of a cluster as an observable. (cf. \citealt{maturi2009anaapptonumcouweapeadet, 2011MNRAS.416.2527M}). Furthermore the physical assumptions entering the model are minimized since the theoretical prediction is directly linked to observational data. Counting detections above different thresholds allows to construct an observable similar to the mass function. However, the $S/N$-thresholds define detections in the first place. Therefore measurement errors are mainly due to finite number of detections. Second, by focusing on the highest peaks only the most prominent weak lensing peaks are important, which are defined very well. This implies that the observable is very well defined and its distribution is also known directly from the model. Thus any assumption about the distribution can be omitted

We investigate the impact of changes in the cosmological parameters and compare the resulting distributions are compared to numerical $N$-body simulations carried out by \citet{borgani2004X-rprogalclugrocoshydsim} using \textsc{Gadget}-2 (\citet{springel2005cossimcodGAD}.  We then investigate the constraining power (with respect to the matter density $\Omega_\text{m}$, the amplitude of the matter power spectrum $\sigma_8$ and the equation of state parameter $w_0$ for spatially flat cosmologies) of this method by performing a maximum-likelihood analysis on mock data with the characteristics expected for future space-based surveys, such as for examples the Euclid\footnote{\url{http://sci.esa.int/euclid}} ESA space mission \citet{Laureijs11}, for which we assume a sky coverage of $15.000\, \text{deg}^2$ and a galaxy number density of $40\,\text{arcmin}^{-2}$.

Throughout this work we use the cosmological base parameters from \citet{collaboration2013Pla201resXVICospar}, namely the matter density parameter $\Omega_\mathrm{m} = 0.314$, the cosmological constant density parameter $\Omega_{\Lambda} = 1 - \Omega_\mathrm{m}$, the baryon density parameter $\Omega_\mathrm{b0}h^2 =0.02206$, the Hubble constant $h= 0.686$ and the normalization of the matter power spectrum $\sigma_8 = 0.834$ as fiducial values.

The structure of the paper is as follows: in Sect. \ref{sec:2} we summarise the theoretical model used for the abundance of cosmic shear peaks describing the different contributions. Further the embedding of the detections due to clusters into the LSS is taken into account in an extended model. In Sect. \ref{sec:3} GEV and order statistics are introduced briefly as well as their connection to cosmology. The model predictions are then compared to numerical $N$-body simulations in Sect. \ref{sec:4}. Also the model presented in \citet{2011MNRAS.416.2527M} is compared to the simulation. In Sect. \ref{sec:5} we discuss the optimal splitting of a wide field survey into sub-surveys in order to project out the desired objects, i.e. clusters. The constraints on cosmology are finally discussed in Sect. \ref{sec:6}.

\section{Modelling Number Counts}\label{sec:2}
In this section, we provide an analytical recipe to predict the number of peaks in weak lensing maps (For a complete review on weak lensing basics see \citealt{2001PhR...340..291B}). The model includes the contribution from observational noise (such as shot and shape noise), large scale structures and non-linear structures such as galaxy clusters. The advantage of having a model predicting all contributions at once is that it avoids the difficult task of splitting these components and keep those arising from galaxy clusters only with the risk of having a sample contaminated by spurious detections and biased constraints of the mass function.

\subsection{Weak gravitational lensing}
The properties of an isolated lens are given by its lensing potential
\begin{equation}\label{eq:1}
\psi(\boldsymbol{\theta}) =\frac{2}{c^2}\frac{D_\mathrm{ds}}{D_\mathrm{d}D_s}\int\limits_0^\mathrm{s}\Phi(D_\mathrm{d}
\boldsymbol{\theta},z)\mathrm{d}z,
\end{equation}
where $\Phi$ is the Newtonian potential and $D_\mathrm{s}$, $D_\mathrm{d}$ and $D_\mathrm{ds}$ are the angular diameter distances between observer and source, observer and lens and lens and source respectively. The integral is carried out along the line-of-sight, with $s$ being the physical distance to the source. $\psi$ relates angular positions of the source $\boldsymbol{\beta}$ to positions of its images $\boldsymbol{\theta}$ via the lens equation $\boldsymbol{\beta}=\boldsymbol{\theta}-\boldsymbol{\alpha}$, where $\boldsymbol{\alpha}= \boldsymbol{\nabla}\psi$ is the deflection angle. For small deflections, the lens equation can be locally linearised, introducing a linear mapping described by the Jacobian
\begin{equation}\label{eq:2}
\mathcal{A}=(1-\kappa)\begin{pmatrix}
1-g_1 && -g_2 \\
-g_2 && 1+g_1\\
\end{pmatrix},
\end{equation}
where $g_i \equiv \gamma_i/(1-\kappa)$ is the reduced shear and $\kappa=\frac{1}{2}\Delta\psi$ is the convergence. The factor $(1-\kappa)$ describes the isotropic magnification of the image, while the complex shear, $g$, describes its distortions. Only the reduced shear $g$ can be obtained by measurements, as the size of the source is unknown. However, in the weak lensing regime, for which $\kappa \ll 1$, $\gamma \approx g$.

\subsection{Lensing signal}
Weak lensing peak counts can be described as the sum of three contributions:

i) The signal due to real objects, i.e. galaxy clusters. Their intrinsic abundance is given by the mass function (\citealt{press1974Forgalclugalbyselgracon, jenkins2001masfundarmathal, sheth1999Larbiapeabacspl}) and the expected density profile.

ii) The signal due to chance projections of the LSS which is described by the projected two-dimensional power spectrum is given by (\citealt{1953ApJ117134L})
\begin{equation}\label{eq:3}
P_\kappa(\ell)=\frac{9H_0^4\Omega_{\text{m}}^2}{4c^4}\int\limits_0^{w_\text{H}}\text{d}w\frac{\bar{W}^2(w)}{a^2(w)}P_\delta\left(\frac{\ell}{f_K(w)},w\right),
\end{equation}
where $w_\text{H}$ is the comoving distance to the horizon and $\bar{W}(w)$ is a weight function including the line of sight integral over the distribution of background sources $G(w)$,
\begin{equation}\label{eq:4}
\bar{W}(w)=\int\limits_w^{w_\text{H}}\text{d}w'G(w')\frac{f_K(w'-w)}{f_K(w')}.
\end{equation}
In this work the distribution $G(w)$ is taken from \citet{benjamin2007Coscon100weasur} with parameters $(a,b,c) = (0.748,3.932,0.8)$. Note that this parameter choice implies a distribution with mean redshift $\langle z\rangle \approx 1$.
As Eq. (\ref{eq:3}) gives the total convergence, the tangential component used in (\ref{eq:7}) is $P_{\gamma_t}=P_\kappa /2$ due to isotropy.

iii) The observational noise is given by the intrinsic ellipticity and the finite number of galaxies used to measure the shear. This contribution has a white power spectrum:
\begin{equation}\label{eq:5}
P_\epsilon =\frac{\sigma^2_{\epsilon_\text{s}}}{2n_\text{g}},
\end{equation}
with the number density of source galaxies $n_\text{g}=40\,\text{arcmin}^{-2}$  and ellipticity variance $\sigma^2_{\epsilon_\text{s}}=0.3$ (When comparison with the ray-tracing simulation $n_\text{g}=30\,\text{arcmin}^{-2}$ and $\sigma^2_{\epsilon_\text{s}}=0.25$ are used.)  

As the contributions by LSS and the observational noise are described by Gaussian random fields, the total contribution to the signal is given by the sum of their respective power spectra,
\begin{equation}\label{eq:6}
P(\ell) =P_{\gamma_t}(\ell)+P_\epsilon.
\end{equation}

\subsection{Optimal filtering}\label{subsec:2.3}
A method to measure the signal of weak gravitational lensing is the aperture mass (\citealt{1996MNRAS283837S}), which is a weighted average of the tangential shear at the position $\boldsymbol{\theta}$:
\begin{equation}\label{eq:7}
A(\boldsymbol{\theta}) = \int_{\mathbb{R}_2}\text{d}^2\theta'\gamma_\text{t}(\boldsymbol{\theta'},\boldsymbol{\theta})Q(|\boldsymbol{\theta'}-\boldsymbol{\theta}|).
\end{equation}
Since we are interested in the signal of non-linear structures and therefore in the exponential tail of the mass function this filter should maximize the \text{$S/N$}-ratio of non-linear structures. A linear filter constructed for this purpose was introduced by \citet{2005A&A442851M} and reads in Fourier space:
\begin{equation}\label{eq:8}
\hat{Q}(\boldsymbol{\ell})=\alpha\frac{\hat{\tau}(\boldsymbol{\ell})}{P(\ell)}, \ \ \ \text{where} \ \ \ \alpha^{-1} = \int_{\mathbb{R}_2}\text{d}^2\ell\frac{|\hat{\tau}(\boldsymbol{\ell})|^2}{P(\ell)}.
\end{equation}
Here $P(\ell)$ is given by equation (\ref{eq:6}) and $\hat{\tau}$ is the expected shear profile of halos. here we assume NFW halos (\citealt{navarro1996UniDenProHieClu}) for which the shear profile is given analytically (\citealt{bartelmann1996Arcunidarhalpro}). The variance of the aperture mass estimate in polar coordinates is
\begin{equation}\label{eq:9}
\sigma^2_A\equiv\int\limits_0^\infty\frac{\ell\text{d}\ell}{2\pi}P_\epsilon|\hat{Q}(\ell)|^2.
\end{equation} 
When applied to ellipticity catalogues of galaxies the aperture mass (\ref{eq:7}) can be approximated by
\begin{equation}\label{eq:10}
A(\boldsymbol{\theta})=\frac{1}{n}\sum_{i=1}^n\epsilon_{t,i}(\boldsymbol{\theta})Q(|\boldsymbol{\theta}_i-\boldsymbol{\theta}|),
\end{equation}
where $n$ is the number density of galaxies and $\epsilon_{t,i}$ is the tangential ellipticity of the $i$-th galaxy with respect to the position $\boldsymbol{\theta}$ which provides an estimate for the tangential shear. $Q$ is the filter function in real space given by the Fourier transform of Eq. (\ref{eq:8}).

\subsection{Number counts for linear and non-linear structures}
The abundance of weak lensing peaks can be described by:

(1) The contribution by non-linear structures which depends on the mass function and their shear profile. Due to shot noise from the discrete positios of background galaxies and their intrinsic ellipticities, its contribution to the aperture mass $A$ given by Eq. (\ref{eq:10}) will scatter around its expectation value $\hat{A}(M)$, where $M$ is the halo mass, with the variance given by Eq. (\ref{eq:9}). Thus halos with mass $M$ will not have a unique signal amplitude, but a probability to produce a certain signal $p(A|M)$, which can be modeled as a Gaussian with width $\sigma_A$. It has been shown by \citet{bartelmann2002Halconweanumcoudarenecos} that the probability for the \text{$S/N$}-ratio$\,\equiv\mathcal S$ to exceed some threshold $\mathcal S_\text{th}$ is given by
\begin{equation}\label{eq:11}
p\left(\mathcal S>\mathcal S_\text{th}|M,z\right)=\frac{1}{2}\text{erfc}\left[\frac{\mathcal S(M,z)-\mathcal S_\text{th}}{\sqrt{2}}\right].
\end{equation} 
Thus, the number of lenses in a redshift interval d$z$ and mass interval d$M$ is given by
\begin{equation}\label{eq:12}
\frac{\text{d}^2n_\text{nl}(\mathcal S>\mathcal S_\text{th},M,z)}{\text{d}z\text{d}M}=p\left(\mathcal S>\mathcal S_\text{th}|M,z\right)n(M,z),
\end{equation}
where $n(M,z)$ is the mass function (in this work we present the difference between the \citealt{press1974Forgalclugalbyselgracon}, \citealt{jenkins2001masfundarmathal} and \citealt{sheth1999Larbiapeabacspl} mass functions). Integrating over the volume d$V$ and d$M$ yields the total number of weak lensing peaks caused by clusters (see \citealt{2011MNRAS.416.2527M}).

(2)
The contribution by the LSS and the instrumental noise which can both be described by a Gaussian random field. In fact, the convergence field is not perfectly Gaussian, however if the applied filter is chosen to be large enough, only small deviations from Gaussianity remain (\citealt{maturi2009anaapptonumcouweapeadet}) 

The number density of detections above some signal threshold $S = {S/N}\sigma_A\equiv x\sigma_A$ is given by
\begin{equation}\label{eq:13}
n_\text{lin}(>S)=\frac{1}{4\sqrt{2}\pi^{3/2}}\left(\frac{\sigma_1}{\sigma_0}\right)^2 \frac{S}{\sigma_0}\exp\left(-\frac{S^2}{2\sigma_0^2}\right),
\end{equation}
where $\sigma_j^2$ are the spectral moments generated by the LSS and the noise:
\begin{equation}\label{eq:14}
\sigma_j^2 = \int\limits_0^\infty\frac{\ell^{2j+1}\text{d}\ell}{2\pi}P(\ell)|\hat{Q}(\ell)|^2.
\end{equation}

Finally the total number density of detections is given as the sum 
\begin{equation}\label{eq:15}
n_\text{det}(>x) = n_\text{nl}(>x) + n_\text{lin}(>x).
\end{equation} 
In this work we distinguish between negative and positive detections. Positive peaks are caused by over-densities and thus also by non-linear objects, hence both terms of Eq. (\ref{eq:15}) have to be taken into account. In contrast the negative detections evolve linearly only and are therefore described by the LSS term only. Note again that we use the definition of detections (\citealt{maturi2009anaapptonumcouweapeadet}) rather than peaks. 
A detection is thus given by a continuous region in a lensing catalogue above a given threshold.
However, for our application these definitions tend to coincide, as we are only interested in the highest peaks.

\subsection{Correction term}
The theoretical model for the positive detections according to Eq. (\ref{eq:15}) does not reproduce the abundance found in $N$-body simulations (see \autoref{Fig:5} and Sect. \ref{sec:4}), because the two contributions entering in Eq. (\ref{eq:15}) are not independent. In fact Eq. (\ref{eq:12}) implicitly assumes that the clusters are placed on the mean signal (which is zero) of the background caused by the LSS.
However, the LSS randomly shifts the expected signal for a halo around its mean value. We account for this effect by thinking of $n_\text{nl}$ as a cumulative distribution, i.e. the number of objects above a certain threshold. This random shift will move peaks from lower to higher signals and reverse with equal probability. However, as there are more objects at a lower signal, this shift will effectively increase the number of objects at higher signals. To see this, consider some ${S/N}$ threshold $({S/N})_\text{th}$ and a Gaussian PDF $p(x,\tilde\sigma_0)$ with zero mean and width 
\begin{equation}\label{eq:16}
\tilde\sigma_0^2 =\frac{\sigma_0^2}{\sigma_A^2},
\end{equation} 
\begin{figure}
\includegraphics[width=0.45\textwidth]{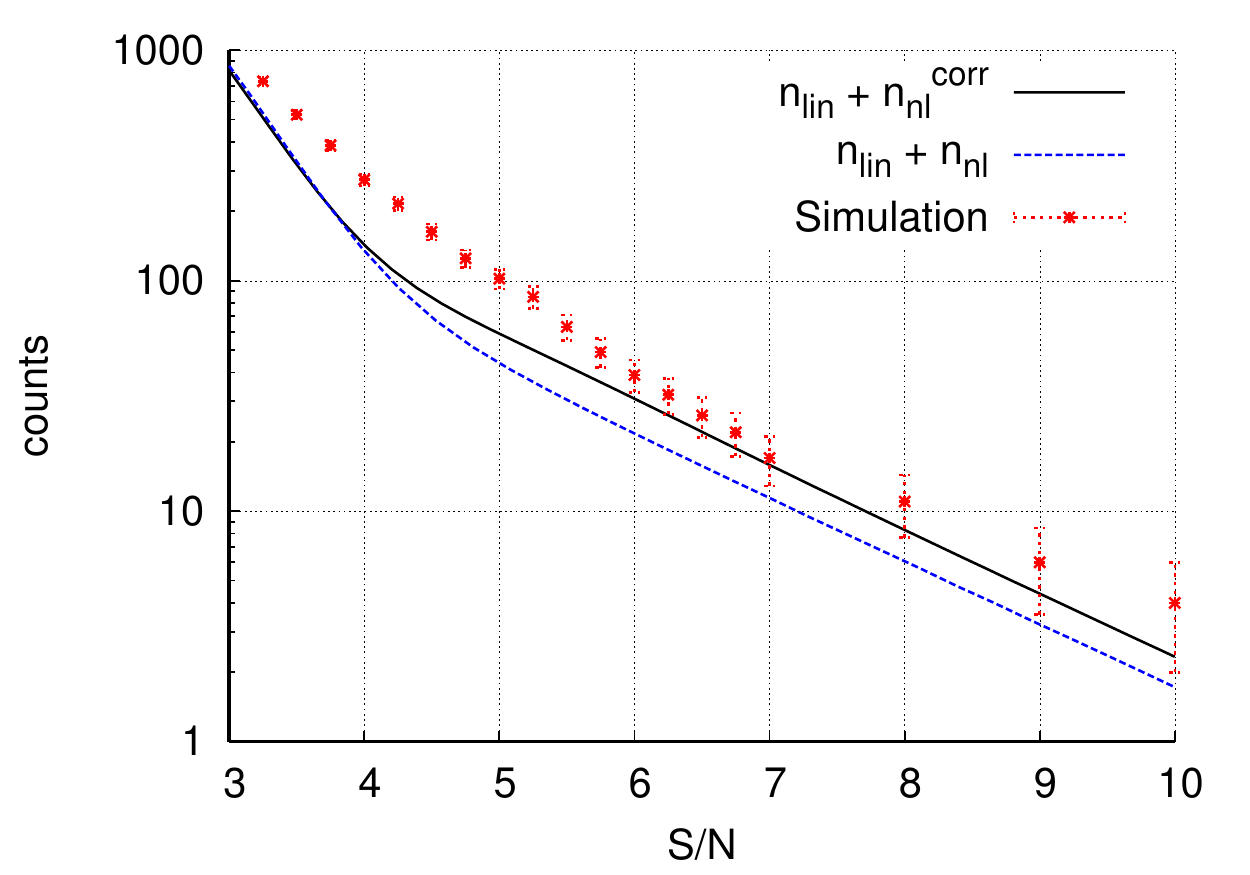}
\includegraphics[width=0.45\textwidth]{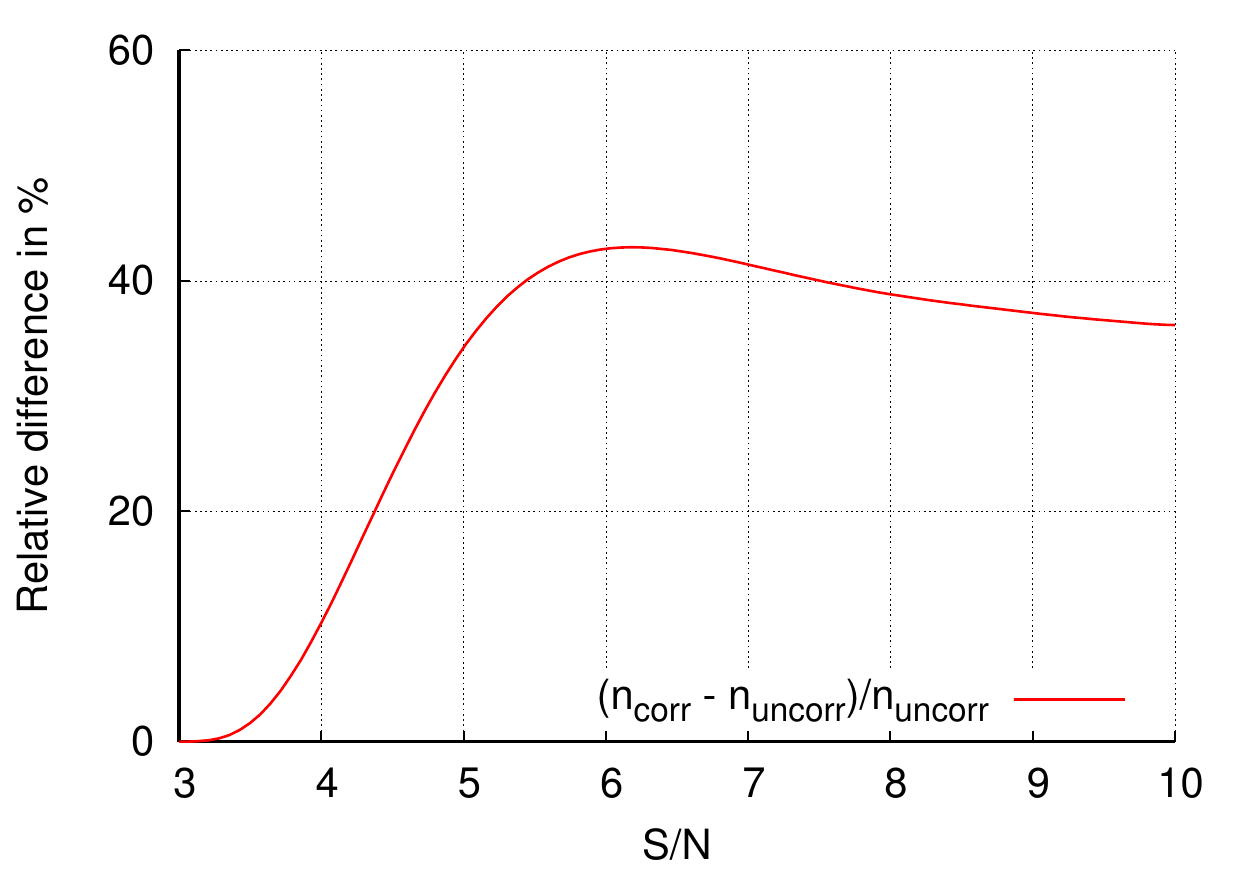}
\caption[Impact of the correction to the non-linear counts]{\textit{Top}: Comparison of the total counts above some \text{$S/N$}-ratio between model and simulation. The blue dotted curve shows the uncorrected model $n_\text{nl}+n_\text{lin}$, while the black solid curve shows $n_\text{nl}^\text{corr}+n_\text{lin}$ from Eq. (\ref{eq:18}). \textit{Bottom}: The relative impact of the correction as a function of \text{$S/N$}.}
\label{Fig:1}
\end{figure}where $\sigma_0^2$ is the variance due to the LSS together with the noise contribution and $\sigma_A$ is the variance of the aperture mass due to the noise. At a certain ${S/N}$ a small fraction of detections moves randomly around its expected value to larger or smaller values. Therefore the corrected number counts are, again with ${S/N}=x$,
\begin{equation}\label{eq:17}
n_\text{nl}^\text{corr} (>x) = n_\text{nl}(>x)+\text{gain}(x)-\text{loss}(x),
\end{equation} 
where loss$(x)$ is the number of detections which are removed from the $S/N$ bin because their amplitude became too large or small due to the scatter, while gain$(x)$ is the number of detections which are moved from other bins into the considered bin. Thus, more explicitly we have:
\begin{equation}\begin{split}\label{eq:18}
n_\text{nl}^\text{corr} (>x) = n_\text{nl}(>x)- & \int\limits_{x_\text{min}}^x \text{d}\tilde x n'_\text{nl}(>\tilde x)p(\tilde x-x;\tilde\sigma_0) \\
+&\int\limits_{x}^\infty \text{d}\tilde x n'_\text{nl}(>\tilde x)p(\tilde x-x;\tilde\sigma_0),
\end{split}\end{equation}
where the prime denotes the derivative with respect to $\tilde x$. The additional minus sign is due to the fact that the number density at the threshold $x$ is needed, which is given by the negative derivative of $n_\text{nl}(>x)$.
The symmetry is now broken due to the fact that $n_\text{nl}$ is monotonically decreasing.
In the upper panel of \autoref{Fig:1} we compare the corrected model (black line, Eq. (\ref{eq:13})+(\ref{eq:18})) and the uncorrected (blue line, Eq. (\ref{eq:15})) mode with a raytracing simulation (see sect. \ref{sec:4} for details). 

The bottom panel of \autoref{Fig:1} shows the variation due to the correction. The increase is roughly 40$\%$ for ${S/N}\gtrsim  6$, but negligible for ${S/N} <3$, where spurious detections due to LSS and noise contribute strongly to the WL field, thus the impact of the correction is small as it does not influence the linear regime. Altogether the correction is small at low $S/N$ but becomes large in the relevant $S/N$ ranges.

We finally note remaining discrepancies. Especially for ${S/N}\in [3,5]$, which is the regime where the contribution of non-linear and linear counts are comparable. In this regime the detections caused by the linear contribution and the non-linear structures will start to blend affecting the statistics in a way which cannot be modeled analytically. A possible way out would be the creation of a mock WL map based on a Gaussian random fields with power spectrum $P(k)$ given by Eq. (\ref{eq:3}) and adding halos according to their mass function and signal profile. A similar approach has been followed by \citet{2014arXiv1410.6955L}. However, this regime is not of interest for our study as we consider only the peaks with extreme values (with large positive $S/N$) far away from this regime. Additionally we are interested in an analytic prescription to evaluate the extreme value statistics. 

Furthermore we neglected the influence of halo triaxiality (see \citealt{2009MNRAS.392..930O} and \citealt{2012MNRAS.425.2287H}). This effect will introduce an additional scatter into the detection amplitude.

\section{General Extreme Value and Order Statistics}\label{sec:3}
This section will review the basic concepts of general extreme value (GEV) statistics and order statistics and how they can be calculated from the fiducial distribution given by the weak lensing counts given by Eq. (\ref{eq:15}) and (\ref{eq:18}).

\begin{figure}
\includegraphics[width = .45\textwidth]{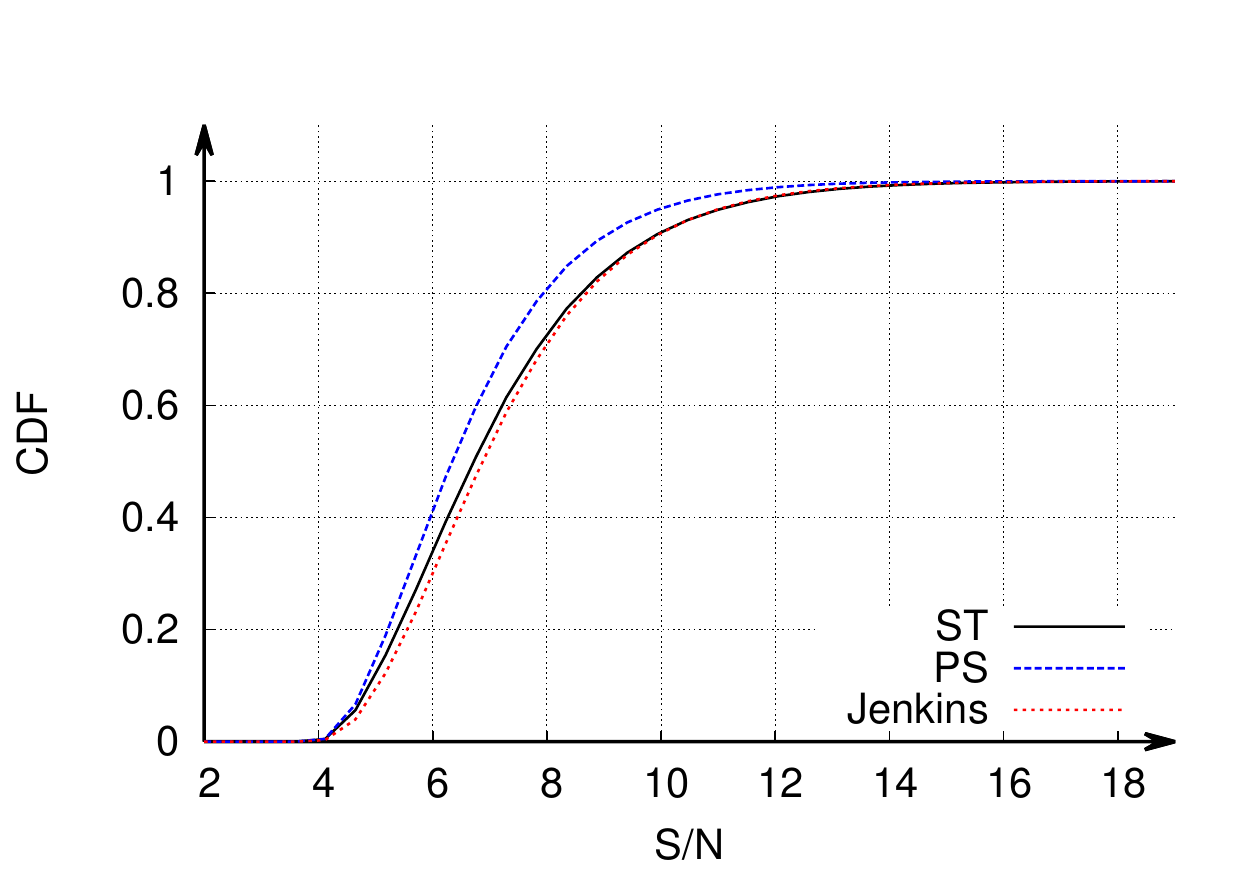}
\caption{The CDF according to Eq. (\ref{eq:19}) of the highest positive detections assuming a survey area $A=1\, \text{deg}^2$ for different mass functions, namely the Sheth-Tormen (solid black line), Jenkins (dashed red line) and the Press-Schechter mass functions (dashed blue line).}
\label{Fig:2}
\end{figure}
\begin{figure}
\includegraphics[width = .45\textwidth]{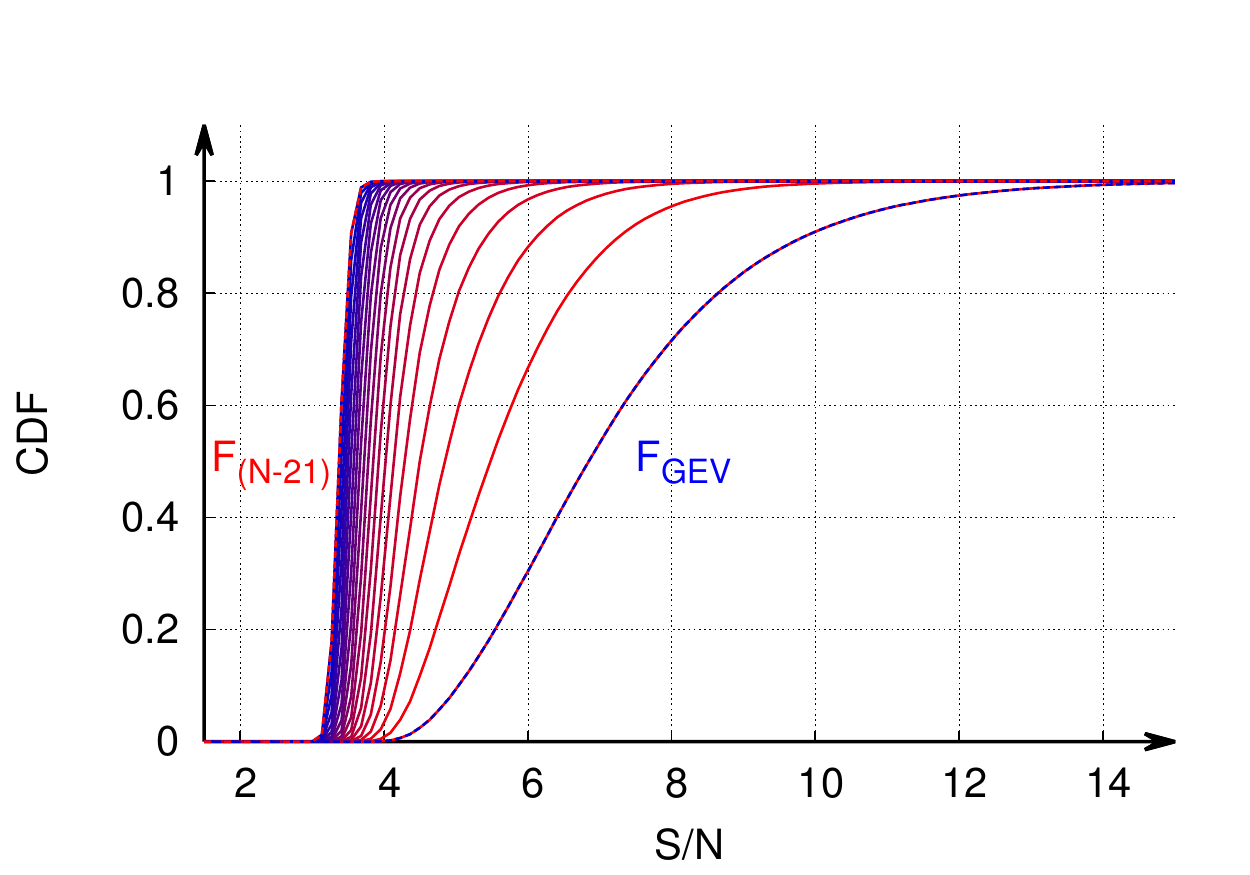}
\caption{The CDFs of the first 22 orders, i.e. $F_{(N-21)}$ to $F_{(N)}$, of the positive detections for a survey area of $A=1\, \text{deg}^2$. Also shown is the corresponding GEV distribution as a dashed line.}
\label{Fig:3}
\end{figure}

\subsection{General extreme value statistics}\label{subsec:31}
The statistics of the maxima and minima of independent and identically distributed random variables is called extreme value statistics (\citealt{gumbel2012statistics}). Consider independent and identically distributed random variables $X_i$ and their block maximum $M_n=\text{max}(X_1,...,X_n)$. As shown by \citet{jenkinson1955fredisannmaxorminvalmetele} and others, the limiting cumulative distribution function (CDF) of $M_n\equiv x$ for $n\to\infty$ is given by
\begin{equation}\label{eq:19}
P_{\gamma,\alpha,\beta}(x) =
\left\{ \begin{array}{ll}
\exp\left(-\left[1+\gamma\left(\frac{x-\alpha}{\beta}\right)\right]^{-\frac{1}{\gamma}}\right) & \ \ \ \text{for  $\gamma\ne 0$} \\
\exp\left(\exp\left(-\frac{x-\alpha}{\beta}\right)\right) & \ \ \ \text{for $\gamma = 0$}
\end{array}\right.,
\end{equation}
where $\gamma$ is the shape parameter, while $\alpha$ and $\beta$ are the so called location parameters. $\gamma = 0$ corresponds to a Gumbel distribution, $\gamma<0$ to a Weibull distribution and $\gamma>0$ to a Fr\'{e}chet distribution. The corresponding probability density function (PDF) can be obtained by differentiating the CDF and is denoted with $p_{\gamma,\alpha,\beta}$.  To evaluate the GEV parameters we consider the weak lensing signal ${S/N}\equiv x$ as a random variable. The CDF of having the largest signal is
the probability of finding no weak lensing peak larger than $x$. We denote this probability $P_0(x)$. Neglecting correlations, i.e. the clustering of clusters, (\citealt{2011MNRAS.413.2087D}), $P_0(x)$ is described by a Poisson distribution for zero occurrence. Thus
\begin{equation}\begin{split}\label{eq:20}
P_{\gamma,\alpha,\beta}(x) \equiv \mathrm{prob}(x_\mathrm{max}\le x) &\equiv\int\limits_0 ^x p_{\gamma,\alpha,\beta} (x_\mathrm{max})\mathrm{d}x_\mathrm{max}\\ & 
\overset{!}{=}P_0(x)=\frac{\lambda^k\exp(-\lambda)}{k!} \\ & \overset{k=0}{=}\exp(-\lambda),
\end{split}\end{equation}
where $\lambda$ is the expected number of peaks in some patch of the sky having a signal larger than $x$. Therefore $\lambda = An_\text{det}(x)$, where $A$ is the survey area and $n_\text{det}(x)$ is the number density of detections given by (\ref{eq:15}) or (\ref{eq:18}) for the old and new model respectively. Now the parameters $\alpha$, $\beta$, $\gamma$ from Eq. (\ref{eq:19}) can be derived for the specific peak number density $n_\text{det}(x)$ by applying a Taylor expansion around the maximum of the corresponding PDFs and equating coefficients. This yields 
\begin{equation}\label{eq:21}
\gamma = n(x_0)A-1, \ \  \beta = -\frac{(1+\gamma)^{1+\gamma}}{n'(x_0)A}, \ \  \alpha = x_0-\frac{\beta}{\gamma}(1+\gamma)^{-\gamma}-1,
\end{equation}
where the prime denotes the derivative with respect to $x$ and  $x_0$ is the most likely signal which can be found by running a root search on
\begin{equation}\label{eq:22}
n''(x_0)-A(n'(x_0))^2=0,
\end{equation}
as a result of the constraint $P_0''(x_0)=0$ (see \hyperref[app:1]{appendix A.1} for a derivation). The distribution of the largest weak lensing signals can thus be described by the three GEV parameters $\alpha$, $\beta$, $\gamma$ which can be directly obtained from the number counts. 

\begin{figure*}
\begin{center}
\includegraphics[width=0.9\textwidth]{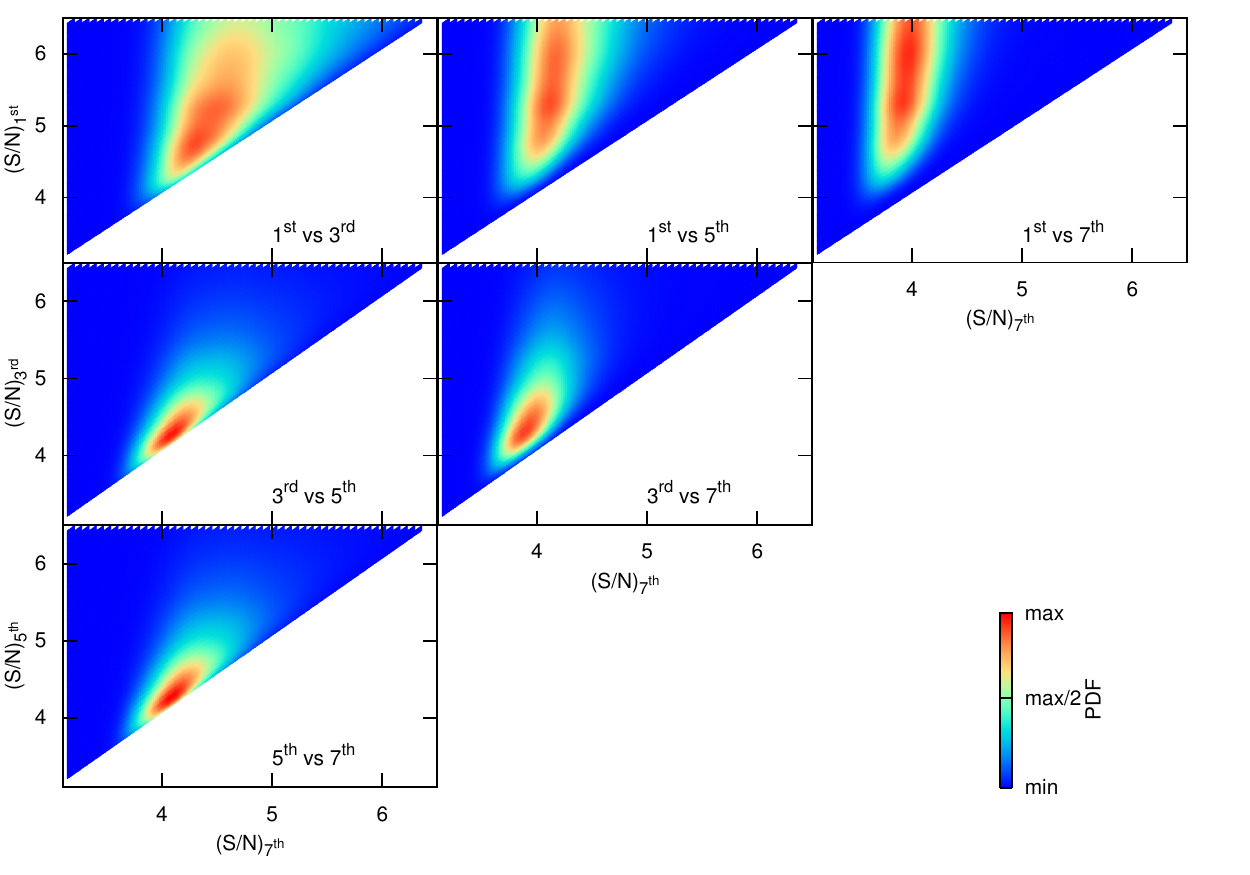}
\caption{Joint PDFs for different combinations of rank orders of the positive peaks. The calculations are carried out with a survey area of 1 deg$^2$. The color ranges from the minimum (blue) to the maximum (red) of the distribution.}
\label{Fig:4}
\end{center}
\end{figure*}

As an example we show in \autoref{Fig:2} the CDFs resulting from Eq. (\ref{eq:20}) for a one square degree survey in the fiducial cosmology, introduced in Sect. \ref{sec:1}, and mass functions: \citet{press1974Forgalclugalbyselgracon}, \citet{jenkins2001masfundarmathal} and \citet{sheth1999Larbiapeabacspl} in blue, red and black, respectively. It can be seen that the choice of the mass function has only a small impact on the distribution of the highest lensing signals except for the PS mass function.

\subsection{Order statistics}
As described in \citet{2013MNRAS.432914W} we now introduce the basic concepts of order statistics, describing the statistics not of the maxima or minima of a distribution, but of their $i$-th highest or lowest values.

Suppose $X_1,X_2,...,X_n$ is a sample of random numbers drawn from some continuous population with a PDF $f(x)$ and a corresponding CDF $F(x)$. The random numbers can now be ordered by magnitude and re-labelled
$ X_{(1)},X_{(2)},...,X_{(n)}$, such that $X_{(i)}\le X_{(i+1)}$. Now, $\forall X_i$ belonging to $x$, i.e. the event $x<X_{(i)}<x+\delta x$ we have a number of $(i-1)$ $X_k$ such that $X_k \le x$. Here exactly one $X_k$ lies in the mentioned interval and the remaining $(n-i)$ $X_k$ have $X_k>x+\delta x$. Since $n$ observations (random numbers) belong to the set, they can be arranged in several ways. The total number of such arrangements is
\begin{equation}\label{eq:23}
A(n,i) =\frac{n!}{(i-1)!(n-i)!}
\end{equation}
with a corresponding probability of having a certain realization
\begin{equation}\label{eq:24}
\mathrm{prob}(x) = [F(x)]^{i-1}[F(x+\delta x)-F(x)][1-F(x)]^{n-i}.
\end{equation}
Here $\delta x$ denotes the width of the interval containing only one element.
For large $n$, $\delta x$ is small and thus terms $\mathcal O (\delta x^2)$ can be neglected yielding
\begin{equation}\begin{split}\label{eq:25}
f_{(i)}(x) &  = \lim_{\delta x\to 0}\left[\frac{\mathrm{prob}\left(x<X_{(i)}\le x+\delta x\right)}{\delta x}\right] \\ & = A(n,i)[F(x)]^{i-1}[1-F(x)]^{n-i}f(x).
\end{split}\end{equation}
Integrating this equation yields the corresponding CDF, which is given by
\begin{equation}\label{eq:26}
F_{(i)} =\sum_{k=i}^n \binom{n}{k}[F(x)]^k[1-F(x)]^{n-k}.
\end{equation}
Thus the distribution of the largest, i.e. the extreme value, or smallest value is
\begin{equation}\label{eq:27}
F_{(n)}(x) = [F(x)]^n \ \ \ \text{and} \ \ \ F_{(1)} = 1-[1-F(x)]^n
\end{equation}
respectively. This means that, given a set of random variables $X_i$, $i\in[1,n]$ drawn from a distribution $F(x)$, the smallest $X_i$ can be described as a random number drawn from $F_{(1)}$, while the largest value is a random variable from $F_{(n)}(x)$.
As discussed in Sect. \ref{subsec:31} for large $n$ both cases can be described by Eq. (\ref{eq:19}). 

This procedure can be extended to joint distributions of several orders. Given two ordered random variables $X_{(r)}$, $X_{(s)}$ such that $1\le r < s \le n$ and $x<y$ their joint PDF is
\begin{equation}\begin{split}\label{eq:28}
f_{(r)(s)}(x,y) & = \ \frac{n!}{(r-1)!(s-r-1)!(n-s)!}\\ &\times [F(x)]^{r-1}[F(y)-F(x)]^{s-r-1}[1-F(y)]^{n-s},
\end{split}\end{equation}
while the joint CDF is given by
\begin{equation}\begin{split}\label{eq:29}
F_{(r)(s)}(x,y)  & = \sum_{j=s}^n\sum_{i =r}^j \frac{n!}{i!(j-i)!(n-j)!} \\ & \times[F(x)]^i[F(y)-F(x)]^{j-i}[1-F(y)]^{n-j}.
\end{split}\end{equation}
\autoref{Fig:3} shows the first 22 orders in a one square degree field for the positive detections. The expected number of observations is therefore $N = n_\text{det}(x_\text{min})A$, with the detection threshold $x_\text{min}$. Note that we need to introduce the cut-off $x_\text{min}$ because $n_\text{det}$ cannot be converted into a distribution function as it is not monotonically decreasing (\citealt{maturi2009anaapptonumcouweapeadet}). However, by setting a boundary, $x_\text{min}$, we restrict ourselves to the region where the distribution is monotonically decreasing. The actual choice of $x_\text{min}$ is not crucial, as long as it converts $n_\text{det}(x)$ into a monotonically decreasing function for all $x>x_\text{min}$ and as long as it is small with respect to the expected values $x$ given by the resulting PDF. We thus write the CDF of weak lensing peak counts as
\begin{equation}\label{eq:30}
F(x) = 1 -\frac{n_\text{det}(x)}{n_\text{det}(x_\text{min})},
\end{equation}
which enters in Eq. (\ref{eq:29}) and (\ref{eq:26}).  
As a first result it can be seen that the CDF steepens with increasing order and thus the constraining power of a single measurement increases. This is due to the fact that for detections with lower $S/N$ more objects occur in the corresponding signal bin. We will call the detection with he highest signal the first order, the detection with the second highest signal the second order and so on.

\autoref{Fig:4} shows the joint probability distribution of different combinations of orders, as denoted in the individual panels. For example we compare the first with the third order in the upper left plot of \autoref{Fig:4} which quantify the probability of finding the largest weak lensing signal at a certain $S/N$, while finding the third largest signal at another $S/N$.  All calculations were done with $A = 1$ deg$^2$ and $x_\text{min} = 1.5$, where $x_\text{min}$ is again the threshold introduced before. 
One can see that the closer the orders are to each other the higher their correlation is. In addition, higher orders confine the PDF to a substantially smaller $S/N$ region.  This reflects the fact that the PDF steepens for higher order combinations, restricting the PDFs to a smaller range in the $S/N$-plane as already shown in \autoref{Fig:3}.

\section{Confronting Theoretical Predictions with Simulations}\label{sec:4}
\subsection{$N$-body numerical simulation}\label{sec:sim}
We test our prediction outlined in Section \ref{sec:2} and \ref{sec:3} by processing a mock ellipticity catalogue derived by a ray-tracing $N$-body numerical simulation carried out with \textsc{GADGET}-2 (\citealt{springel2005cossimcodGAD}) and presented by \citet{borgani2004X-rprogalclugrocoshydsim}. The simulation represents a standard $\Lambda$CDM model including dark matter as well as additional baryonic physics for the gas component modeled by radiative cooling, star formation and supernova feedback in the form of galactic winds. The base cosmological parameters are $\Omega_\text{m} = 0.3$, $\Omega_\Lambda=0.7$ and $\Omega_b = 0.04$. The Hubble constant is $H_0=100 h\text{km}\,\text{s}^{-1}\, \text{Mpc}^{-1}$ with $h = 0.7$ and the linear power spectrum of the matter-density fluctuations is normalised to $\sigma_8 =0.8$. The simulated box has a side length of $192h^{-1}$Mpc, containing $480^3$ dark-matter particles with a mass of $6.6\times 10^9h^{-1}M_\odot$ each and an initially equal number of gas particles with a mass of $8.9\times 10^8 h^{-1} M_\odot$ each.

As discussed in \citet{pace2007Tesrelwealencludet} and \citet{maturi2009anaapptonumcouweapeadet} backward light cones have been constructed using the simulation by taking snapshots at different redshifts ranging from $z=1$ to $z= 0$. In order to avoid repetitions of the same structures in snapshots at different redshifts, these are randomly shifted and rotated. The light cone has been sliced into thick planes whose particles are subsequently projected with a triangular-cloud scheme on lens planes perpendicular to the line-of-sight. Finally a bundle of $2048\times 2048$ light rays is traced through the light cone starting from the observer into directions on a grid of $4.9$ degrees on each side. From this ray tracing simulation effective convergence and shear maps are obtained. These are used to lens a background source population of galaxies at $z=1$ with a number density of $n_g = 30\, \text{arcmin}^{-2}$ and random intrinsic ellipticities drawn from
\begin{equation}\label{eq:31}
p(\epsilon_s) =\frac{\exp[(1-\epsilon_s^2)/\sigma_{\epsilon}^2]}{\pi\sigma_\epsilon^2[\exp(1/\sigma_\epsilon^2)-1]},
\end{equation}  
with $\sigma_\epsilon = 0.25$. The mock catalog of galaxy ellipticities is then analysed with optimal filter described in Sect. \ref{subsec:2.3} on a grid of $512\times 512$ positions covering the entire field-of-view. To account for the survey geometry, the pixelized shear map and the discretization of the sky, a filter functions $W$ needs to be introduced in the spectral moments (Eq.(\ref{eq:14}) when evaluating the number counts model from the simulation. The result of the filtering procedure is 
\begin{equation}\label{eq:32}
\sigma_j^2 = \int \frac{k^{2j+1}\text{d}k}{\pi}P(k)\hat W^2|\hat Q(k)|^2,
\end{equation}
where $Q$ is the optimal filter introduced in Eq. (\ref{eq:8}) and the filter function $W$ is a product of the following terms: a high-pass filter, suppressing scales larger than the side length of the field of view,
\begin{equation}\label{eq:33}
\hat{W}^2_\text{f}(k) = \exp\left(-\frac{k_\text{f}^2}{k^2}\right),
\end{equation}
a low-pass filter, filtering out signatures on scales smaller than the average separation of sources,
\begin{equation}\label{eq:34}
\hat{W}_\text{g}(k) = \exp\left(-\frac{k^2}{k^2_\text{g}}\right)
\end{equation}  
and another low-pass filter including the finite size of the pixels
\begin{equation}\label{eq:35}
\hat{W}_\text{pix} = \frac{2\sqrt{\pi}}{kd_\text{pix}}J_1\left(\frac{kd_\text{pix}}{\sqrt{\pi}}\right),
\end{equation}
where $k_\text{f} =2\pi/L_\text{f}$, $k_\text{g} = 2\pi \sqrt{n_\text{g}}$,$d_\text{pix} = 4.9^\circ/512$ and $J_1$ is a cylindrical Bessel function of order one, representing a circular step function of the size of a pixel.

\begin{figure*}
\begin{center}
\includegraphics[width = .9\textwidth]{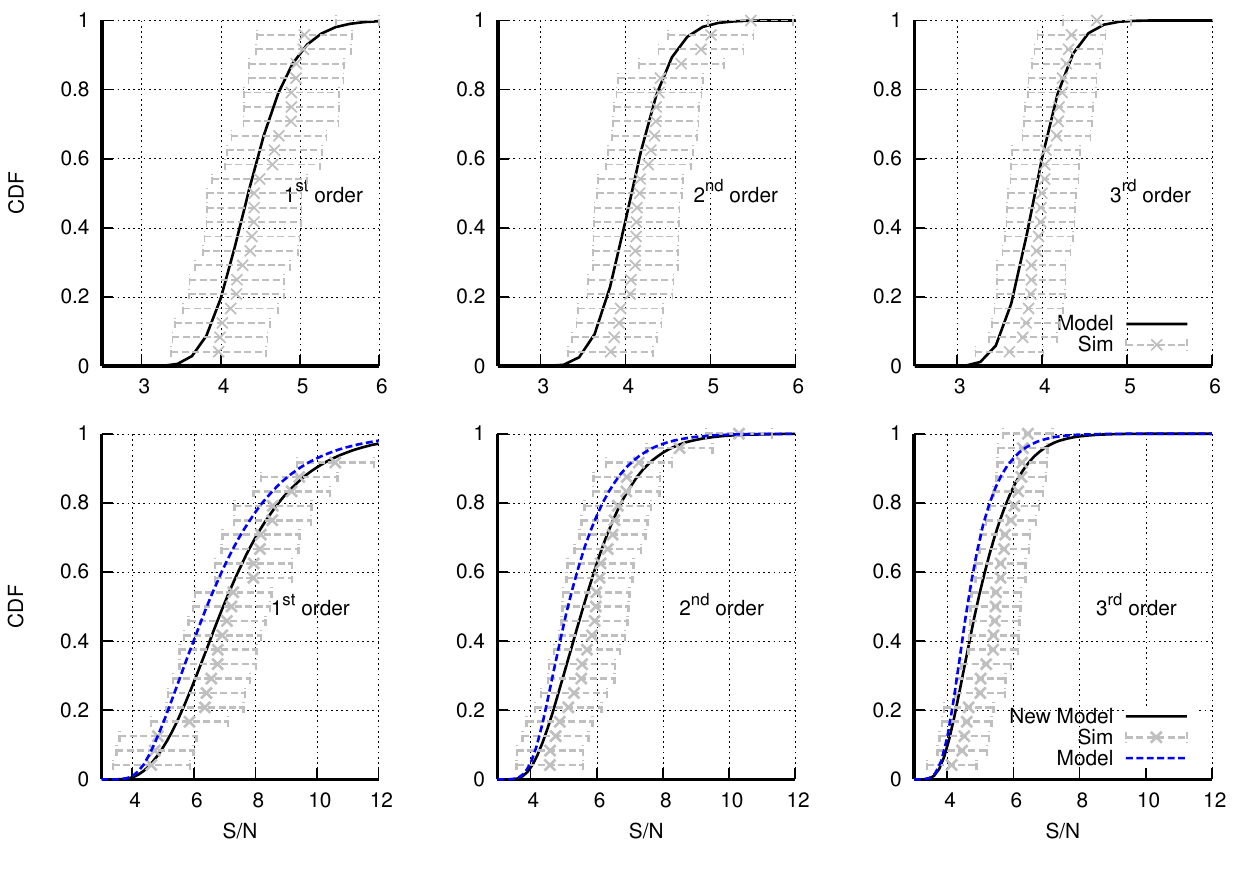} 
\caption{Comparing out analytic predictions with a ray-tracing $N$-body simulation (grey crosses with errorbars). The  survey area, $A$, is $4.9\times 4.9/24 \,\text{deg}^2$ and the cosmological base parameters discussed in Sect. \ref{sec:sim} are used. \textit{Top}: Prediction for the negative detections (black line), i.e. linear structures only (Eq. (\ref{eq:13})). \textit{Bottom}: The model discussed in \citet{2011MNRAS.416.2527M} (see Eq. (\ref{eq:15})) is shown in blue (dashed), while the new model including the correction from Eq. (\ref{eq:18}) is shown in black.}
\label{Fig:5}
\end{center}
\end{figure*}

\subsection{Estimation of the CDF}
The detections are defined by the up-crossing criteria discussed in \citet{maturi2009anaapptonumcouweapeadet}, i.e. the resulting weak lensing maps are sliced into different \text{$S/N$}-ratios. Then the number of detections is counted above every threshold. A detection is defined as a continuous group of pixels above the given threshold, which are linked if they are closer to each other than a certain linking length. In this case the latter is 1.42 pixels, so that also pixels diagonal with respect to each other are linked together.

In order to study the order statistics of the WL maps in the numerical simulation, the field has to be divided into $N$ smaller sub-fields, each with area $A_\text{sub}$. The sample of the $i$-th highest value drawn from each sub-field represents the order statistic of order $i$ corresponding to a survey area $A_\text{sub}$. The CDF of the $i$-th order from the simulation can be estimated by
\begin{equation}\label{eq:36}
F_{(i)}\left(x_{(i)}^j\right) = \frac{1}{N}j, \ \ \ \ j\in[1,N],
\end{equation} 
where $x_{(i)}^j$ is the ordered (from small to large values) vector of the $i$-th highest value in each subfield $A_\text{sub}$. As all $x_{(i)}^j$ are drawn from the same distribution, $f_{(i)}(x)$, their respective error is given by the  width of the distribution.
In Sect. \ref{sec:4.3} and \ref{sec:4.4} we compare the order statistics of both, positive and negative $S/N$, with the simulation. By splitting the WL map into 24 tiles.
Note that the error bars are rather confidence regions which are given by the width $\Delta_{(i)}$ of the PDF of the $i$-th order statistics. We define this to be 
\begin{equation}\label{eq:37}
\Delta_{(i)} \equiv Q_{80} - Q_{20},
\end{equation}
where $Q_p$ is the quantile of the distribution with $p=100\times\text{CDF}(Q_p)$, i.e. the inverse of CDF. The value $x_{(i)}^j$ of the simulation is than given by $x_{(i)}^j = x_{(i)}^j\pm \Delta_{(i)}/2$.

\subsection{Negative detections}\label{sec:4.3}
The first row of \autoref{Fig:5} shows the model prediction for the first three orders together with the distribution resulting from the numerical simulation calculated from Eq. (\ref{eq:36}).  Clearly the prediction lies well within the error bars.
The steepening and shifting to smaller values of the order statistic in the simulation are in good agreement with the theoretical prediction.
We stress the fact that \autoref{Fig:5} shows a prediction and not a fit of the model to the simulated data.

\subsection{Positive detections}\label{sec:4.4}
In the second row of \autoref{Fig:5} we repeat the same exercise for the positive detections. Again, the model is in good agreement with the simulation.
Note that for the first order the CDF from the simulation cannot reach unity because there are two peaks at very high ${S/N}$, namely $S/N = 13$ and $S/N =17$, which are not shown for easier comparison of the different orders.
In this Figure two models are shown: The dashed blue line corresponds to the order statistics calculated from the uncorrected version of the number counts, while the black line includes the correction acting on the detections caused by non-linear structures. This clearly illustrates that the contribution of the LSS is relevant also for the detections with the highest $S/N$ ratios. Ignoring the contribution would lead to biased cosmological parameters, even though the detections at large $S/N$ are only due to non-linear structures. 
We further notice that the difference induced by the correction is much larger than the one due to the use of different mass functions (\autoref{Fig:2}).

\section{Optimizing the area of the sub-fields}\label{sec:5}
The previous analysis has been performed by splitting the simulated field ($4.9\times 4.9\,$ deg$^2$) into $N$ sub-fields. So far the choice was arbitrary and we used a size of $1\,\text{deg}^2$ for simplicity. Thus, the question is whether this can be done in a more objective and optimal way to do it. 

\begin{figure}
\begin{center}
\includegraphics[width = 0.45\textwidth]{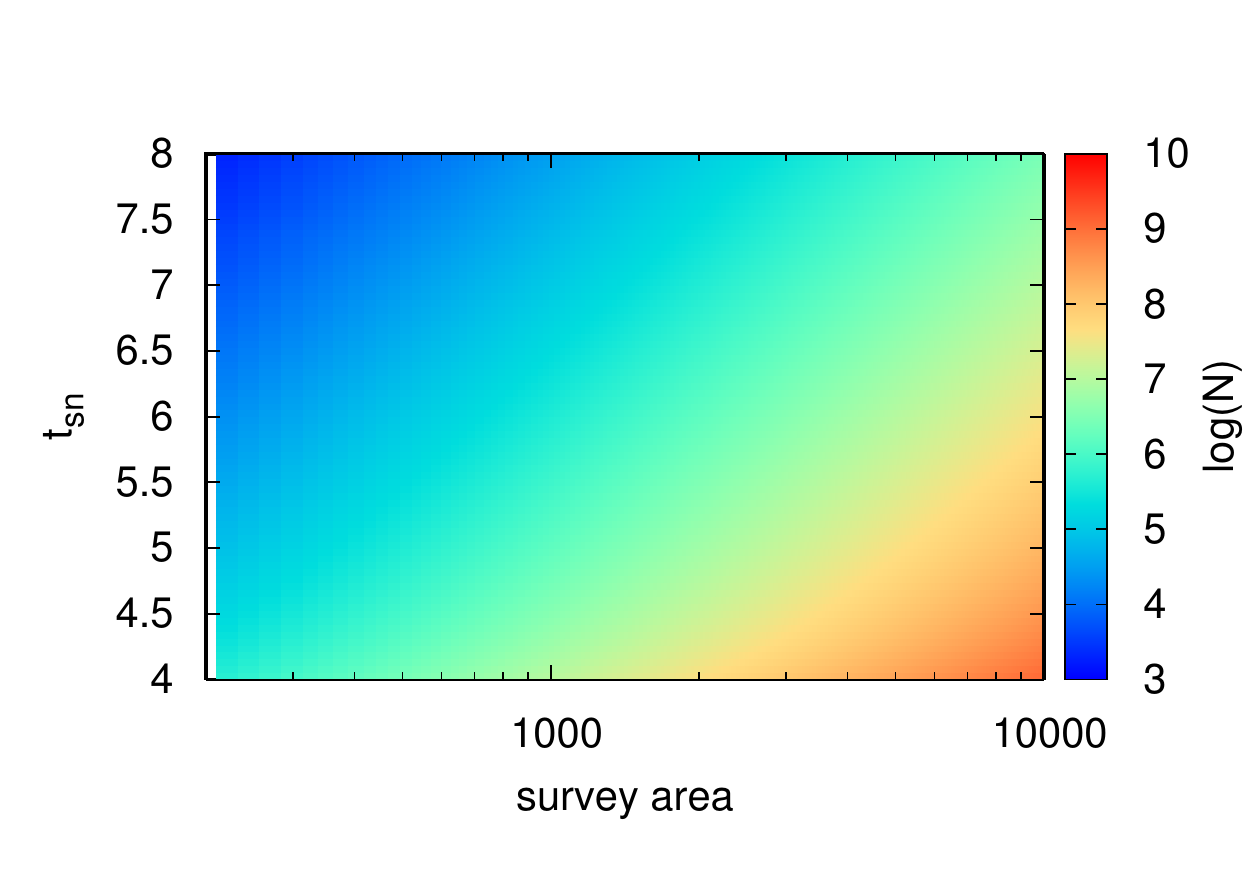}
\caption{Dependence on the optimal number of sub-fields on the threshold $t_{sn}$ and the total survey area.}
\label{Fig:6}
\end{center}
\end{figure}

Since the goal is to estimate some distribution on a large number of sub-fields is preferred as the sampling improves and thus the overall constraining power on cosmological parameters. However, increasing the number of sub-fields automatically decreases the size of each sub-field given a fixed total survey area, $A$. Therefore the strongest lenses in each sub-field will have smaller signals, diluting the presence of those with larger amplitude and mostly related to non-linear structures. Furthermore the sensitivity on cosmological parameters shrinks due to the dilution of non-linear structures and LSS due to the noise.
The probability of finding a signal above some threshold $t_{sn}$ is  
\begin{equation}\label{eq:38}
\text{Prob}\left[t_{sn};N\right] = \int\limits_{t_{sn}}^\infty p_N(x)\text{d}x,
\end{equation}
where $p_\text{N}(x)$ denotes the PDF  of the order statistics corresponding to a sub-field size of $A_\text{sub}=A/N$. Again, increasing $N$ will lead to a better sampling of the distribution, but it will also lead to a lower probability of finding a detection due to a cluster in each sub-field. The expected number of objects above this threshold is thus
\begin{equation}\label{eq:39}
\langle N\rangle = N\text{Prob}\left[t_{sn};N\right].
\end{equation}
The threshold $t_{sn}$ chosen such that the largest signals are likely to be due to clusters. Eq. (\ref{eq:38}) is the probability that a halo with $S/N>t_{sn}$ can be found in a certain area, conversely,
\begin{equation}\label{eq:40}
q \equiv 1-\text{Prob}\left[t_{sn};N\right] \equiv  1-p
\end{equation}
is the probability of not finding a halo. However, aiming at galaxy clusters the largest signal in each sub-field is required to be due to the non-linear contribution. Therefore no object should be below the threshold $t_{sn}$ which marks a boundary between non-linear and linear counts, thus
\begin{equation}\label{eq:41}
qN \le 1,
\end{equation}
such that in each sub-field the corresponding largest signal occurs above the threshold. Eq. (\ref{eq:41}) sets a maximum for $N$. Combining this with the goal to sample the distribution as well as possible the maximum $N$ can be used as the number of sub-fields. As a result we present in \autoref{Fig:6} the number of sub-fields as a function of the threshold $t_{sn}$ and the total survey area. As expected, increasing the threshold lowers the maximum number $N$, since there are fewer objects above higher thresholds. Accordingly with increasing total survey area the $N$ also increases as more data are available. As the total survey area is fixed by the experiment the only free parameter left is $t_{sn}$, which should be chosen such that everything above the threshold will most likely be a halo and not due to LSS. A reasonable value for the threshold is found to be $t_{sn}\approx 5-6$, which is the value at which well observable galaxy clusters show themselves.
\begin{figure}
\begin{center}
\includegraphics[width = 0.45\textwidth]{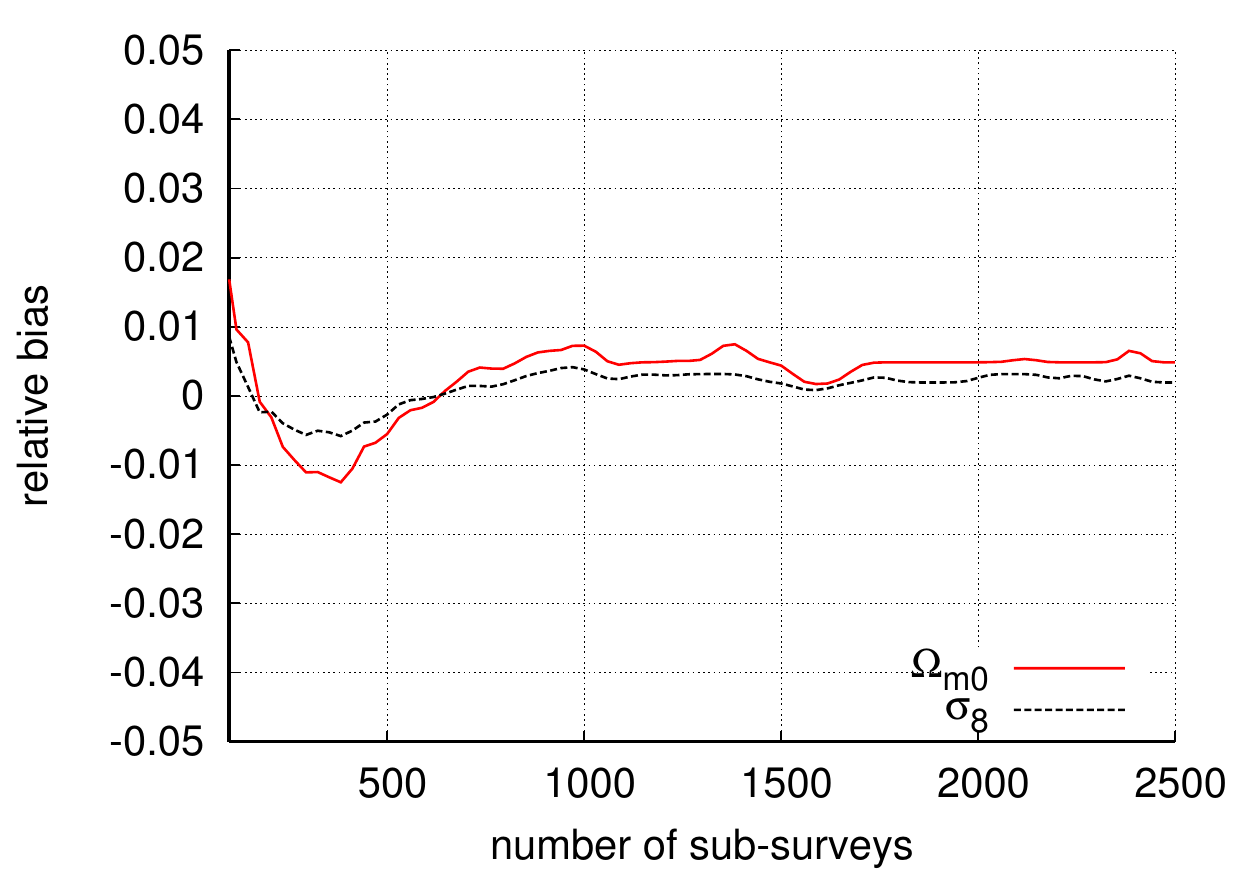}
\caption{The relative sample bias as a function of the number of sub-fields for a fixed sub-field size $A$. $\Omega_\text{m}$ and $\sigma_8$ were fitted separately.}
\label{Fig:10}
\end{center}
\end{figure}
\section{Constraining Cosmology}\label{sec:6}
\begin{figure*}
\begin{center}
\includegraphics[width = 0.9\textwidth]{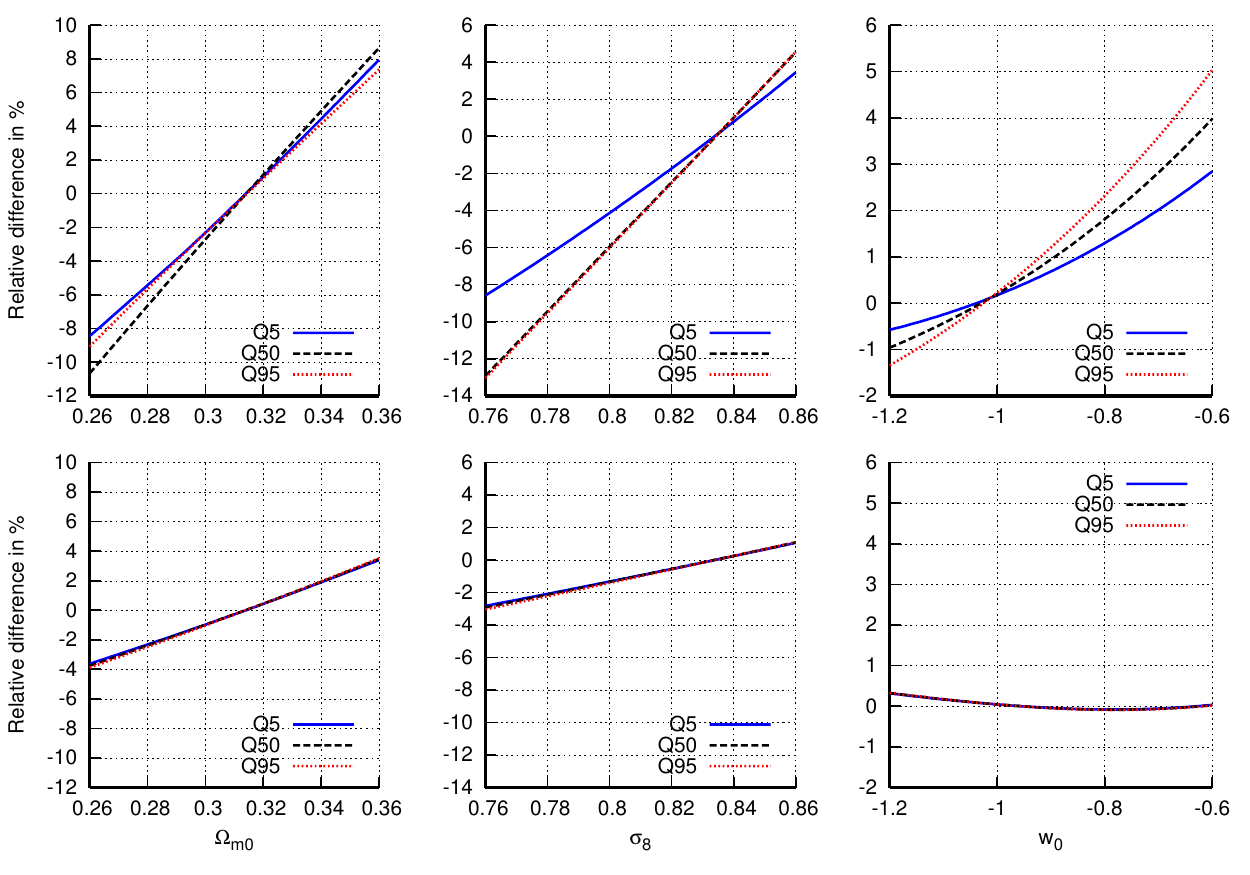}
\caption{Relative deviations $\Delta$ in percent of quantiles of the extreme value distribution (1st order) as a function of cosmological parameters. Quantiles used are $5\%$ (solid blue) $50\%$ (dashed black) and $95\%$ (dotted red). The fiducial model is given by the base parameters described in \autoref{sec:1}. \textit{Top}: positive detections. \textit{Bottom}: negative detections.}
\label{Fig:7}
\end{center}
\end{figure*}
In this section we use the method outlined in Sect. \ref{sec:2} and \ref{sec:3} to constrain cosmological parameters. Before applying the method to mock data we briefly discuss the impact on individual cosmological parameters on the extreme value statistics. Afterwards we carry out a likelihood analysis using a {Euclid}-like survey with $A=15000\,\text{deg}^2$. We mimic a source redshift distribution with a mean redshift of approximately unity, using the parametrisation given in \citet{benjamin2007Coscon100weasur} $(a,b,c) = (0.748,3.932,0.8)$ and a source number density $n_\text{g} = 40\, \text{arcmin}^{-2}$.
We investigate two scenarios, one completely ignoring redshift information and one dividing the survey into two redsift bins. In the first case we use $N = 4574$ sub-fields and in the second case the following number of sub-fields: $N_\text{low} = 2858$ and $N_\text{high} = 2059$ according to the criterion given in Sect. \ref{sec:5}. Mock data are created by sampling $N$ random numbers drawn from the distribution of the largest signal given by Eq. (\ref{eq:22}). Note that small survey sizes, allowing a sampling of $N\sim500$ sub-fields, can create a sample bias when deriving the cosmological parameters because of under-sampling of the fiducial distribution and its non-Gaussianity. This effect can be reduced by using the average likelihood, i.e. by averaging over many realisations of Eq. \ref{eq:44}. This will increase the uncertainties in the derived parameters. In \autoref{Fig:10} the relative bias $(\theta_\text{fit}-\theta_\text{fid})/\theta_\text{fid}$, with $\theta$ being the parameter is shown as a function of the number of sub-fields $N$. It can be seen that the bias increases at low $N$. At higher $N$ it is close to zero and approximately constant. The systematic offset just arises from the finite resolution of the likelihood and can be in principle reduced to zero by introducing more grid points in parameter space. 
Note that the plotted parameters were not fitted simultaneously.
However, for our application the number of sub-fields is much larger and this sample bias will not pose any problem.

\subsection{Impact on cosmology}
We take a first look at the dependence on cosmology of the order statistic ignoring any redshift information as a starting point.

\autoref{Fig:7} shows the dependence of quantiles $Q_p$, i.e. the inverse of the CDF at a given value $p$ of the extreme value distribution (1st order) of both positive and negative peaks as a function of cosmological parameters (for example $Q_{50}$ corresponds to $p=0.5$ and therefore $50\%$ probability occurs below $Q_{50}$). We show the relative deviation from the fiducial model 
\begin{equation}\label{eq:42}
\Delta \equiv (Q_p -Q_p^\text{fid})/Q_p^\text{fid},
\end{equation}
and for which we assumed a flat universe.

The sensitivities to $\Omega_\text{m}$ and $\sigma_8$ are comparable and very strong in case of the positive detections. This also shows the degeneracy between the two parameters which is typical for estimates based on weak lensing.
The shape of the distribution changes slightly when varying the parameters as the curves of different quantiles have different slopes. This effect is due to the highly non-linear dependence of the positive detections on those parameters. This behaviour cannot be seen in the bottom row, where the negative detections are shown. Because of the Gaussian nature of the peaks in this case, the overall shape of the extreme value distribution does not vary significantly when changing cosmological parameters. 
We also note that the sensitivity is much stronger in case of the positive detections, causing changes up to $12\% $ in case of $\Omega_\text{m}$ and $\sigma_8$ in the considered region of parameter space, while the negative detections only give rise to changes up to $4\%$.

The dependence on the equation of state parameter is much weaker than the dependence on parameters related to structure formation. Volume effects are less important, compared to the dependence on $\Omega_\text{m}$ and $\sigma_8$.
The increase of the relative difference for smaller $w_0$ (less negative) is  mainly due to the modifications of the exponential tail of the mass function. It can again be seen that the negative detections show a weak sensitivity compared to the positive peaks with respect to $w_0$. This agrees with the fact that volume effects are not the dominating factor.

Note that when adopting a time dependent equation of state $w(a) = w_0 +w_\text{a}(1-a)$, the quantiles nearly stay constant in the parameter range. In \citet{2013MNRAS.432914W} and \citet{pace2010Sphcolmoddarcos} it has already been shown that the most massive objects are only affected at higher redshifts when adopting a simple linear model for $w(a)$. For halos at high redshift the lensing signal shrinks and thus those objects are not captured.

Additionally, the number of background galaxies $n_\text{g}$ shrinks when going to higher redshifts, thus the \text{$S/N$}-ratio drops as the shot noise increases. Thus, even though  the abundance of very massive halos will change (see \citealt{2013MNRAS.432914W}), the most extreme weak lenses are not affected that much as those massive halos appear at higher redshift, where the \text{$S/N$}-ratio is lower in the first place. Therefore the dependence on $w_ \text{a}$ is very weak as its impact becomes important only at higher redshift when keeping $w_0$ constant. Note that the impact of cosmological parameters also depends on the adopted source redshift distribution, for example a higher sensitivity in $w_0$ can be reached if the sources populate higher redshifts.

We finally note that the corresponding GEV parameters $\alpha$, $\beta$ and 
$\gamma$ depend nearly linearly on the cosmological parameters except for $w_0$, allowing the use of simple schemes to quickly evaluate the likelihood.

\subsection{The likelihood}
For calculating the likelihood a mock data set is created by sampling $N$ random numbers from the fiducial model, i.e. from the derivative of Eq. (\ref{eq:20}) $p_{\alpha,\beta,\gamma}(x)$ evaluated at the fiducial cosmology. The components of the data vector $\boldsymbol D$ are denoted $(\boldsymbol D)_i \equiv X_i$.Those $N$ numbers are the highest \text{$S/N$}-ratios in the $N$ sub-fields with area $A = A_\text{tot}/N$, i.e. we fix $A_\text{tot} = 15000\, \text{deg}^2$ and calculate the optimal number of sub-surveys $N$ via the condition (\ref{eq:41}). Having fixed $A_\text{tot}$ and $N$, the mock data is sampled from the resulting distribution (\ref{eq:20}) for both the positive and negative peaks.

\begin{figure}
\begin{center}
\includegraphics[width = 0.45\textwidth]{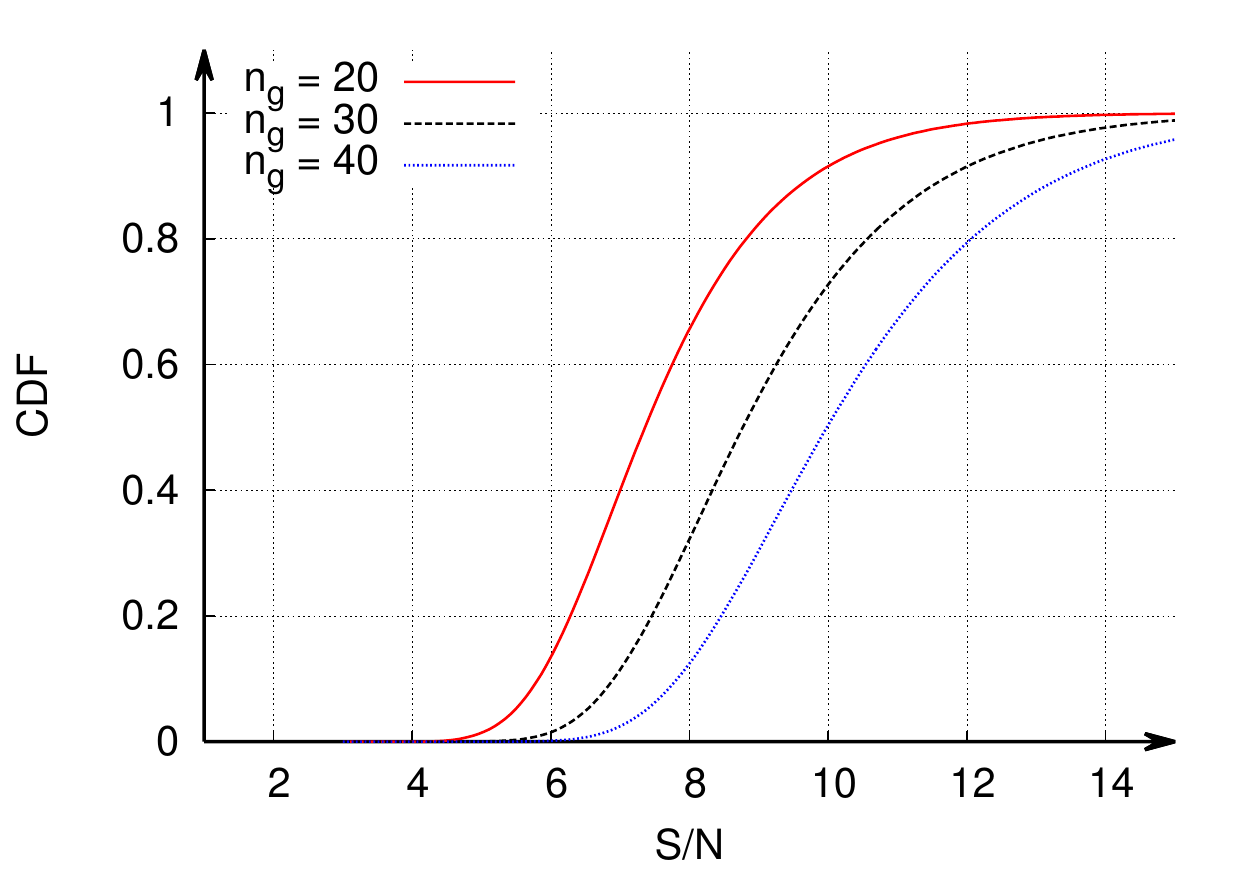}
\caption{Influence of the source number number density $n_g$ on the CDF for fixed area $A$.}
\label{Fig:12}
\end{center}
\end{figure}

The likelihood is now the joint conditional probability for having a set of parameters, collected in the parameter vector $\boldsymbol\Theta$, given a data set $\boldsymbol D$:
\begin{equation}\label{eq:43}
\mathcal L(\boldsymbol\Theta|\boldsymbol D) = \prod _{i=1}^N p_{\alpha,\beta,\gamma,i}(\boldsymbol\Theta|X_i),
\end{equation}
where we already assumed a flat prior $p(\boldsymbol\Theta)$ and the evidence $p(\boldsymbol D)$ was already included into the normalization. The log-likelihood $L$ is
\begin{equation}\label{eq:44}
 L(\boldsymbol\Theta) = \sum_{i = 1}^N\sum_{j = \pm} \log\left( p_{\alpha,\beta,\gamma,i,j}(\boldsymbol\Theta|X_i)\right),
\end{equation}
where the plus and minus signs denote positive and negative peaks respectively. Again flatness is assumed, i.e. $\Omega_\Lambda = 1 - \Omega_\text{m}$, while parameters like $h$ and $\Omega_\text{b}$ are fixed in the first place, as their influence is negligible, thus yielding only an overall factor. Since the individual terms $ p_{\alpha,\beta,\gamma,i,j}$ of the likelihood are not Gaussian or have a simple exponential form, also the likelihood will of course not assume this simple form. Therefore a Fisher approach is not suitable and the likelihood has to be sampled directly or via Monte-Carlo methods.

\begin{figure}
\begin{center}
\includegraphics[width = 0.45\textwidth]{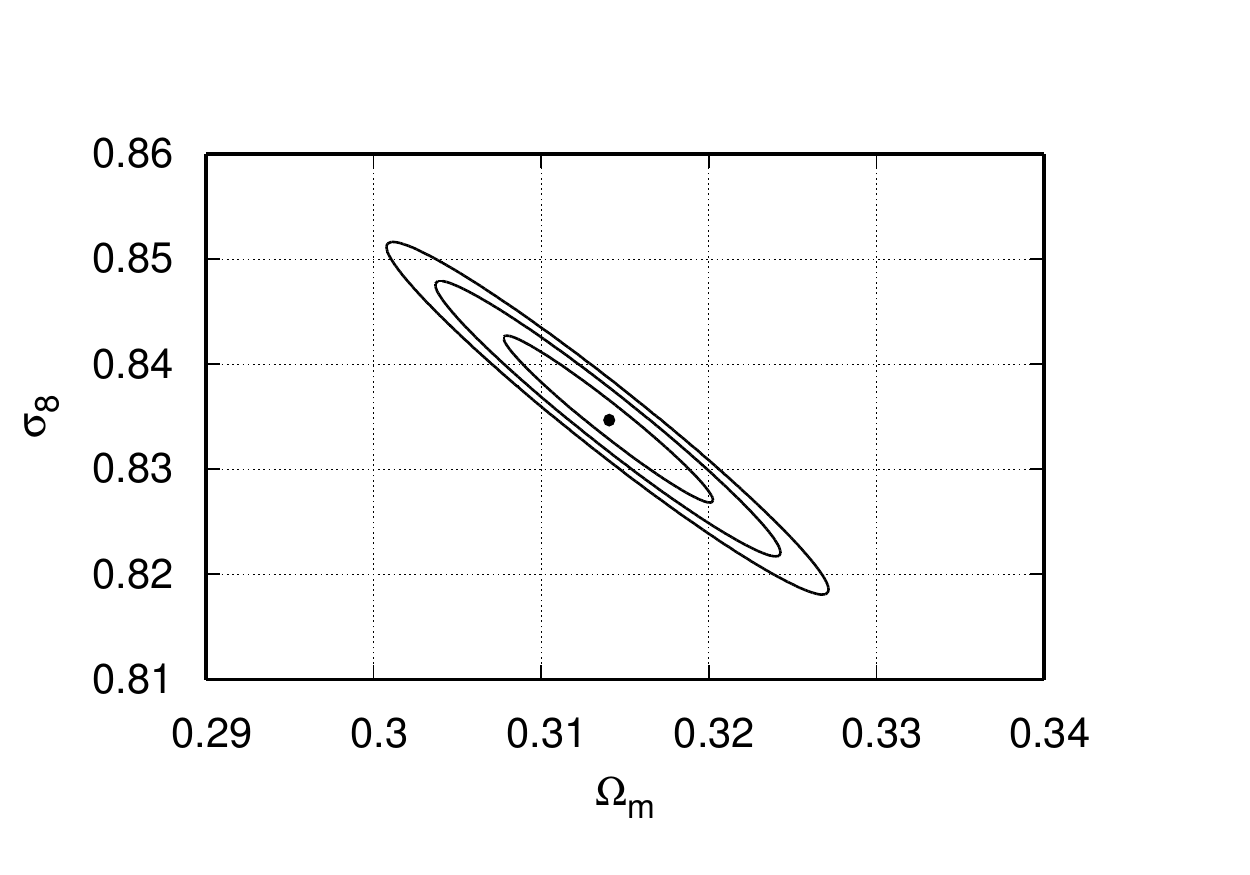}
\caption{Likelihood for a {Euclid} like survey with $A = 15000$ deg$^2$. and without redshift information. We show confidence regions corresponding to $1\sigma$, $2\sigma$ and $3\sigma$. Positive and negative counts are used, i.e. $j=\pm$ in Eq. (\ref{eq:44}). The fit was carried out with $\sigma_8$ and $\Omega_\text{m}$ only. The fiducial cosmology is marked with a dot.}
\label{Fig:8}
\end{center}
\end{figure}

In \autoref{Fig:12} we show the influence of the source number density $n_g$. Clearly, fewer background galaxies lead to a lower weak lensing $S/N$-ratio, which is expected. However, the sensitivity of the highest peaks to cosmological parameters is due to the mass function. Thus the shape of the likelihood will be conserved. Nonetheless, as the $S/N$ decreases if $n_g$ is decreased, the expected $S/N$ of the highest peak will also decrease if the sub-field area is kept fixed. For example, if we are given two surveys with the same total survey area $A_\text{tot}$ but different source densities $n_g$, the possible number of sub-fields $N$ will also differ. Especially if $n_{g1}>n_{g2}$ we have $N_1>N_2$. As the sub-fields are considered to be independent, the likelihood scales with $N^{-1/2}$ due to the Poissonian nature of the statistical error. In the survey discussed in this work the uncertainties of the cosmological parameters for $n_g=40\, \text{deg}^{-2}$ will pick up a factor $\sqrt{N(40)/N(30)}\approx 1.2$ if we calculate the likelihood with $ n_g=30\, \text{deg}^{-2}$. 

When including the two redshift bins, their individual contributions are included in Eq. (\ref{eq:44}) as an additional sum over all redshift bins. Furthermore the upper limit of the summation $N$ is bin-dependent. The likelihood therefore reads $L_z(\boldsymbol\Theta) = \sum L_\text{bin}(\boldsymbol\Theta)$, where $ L_\text{bin}(\boldsymbol\Theta)$ is given by Eq. (\ref{eq:44}).

\subsection{Confidence regions}
\begin{figure*}
\begin{center}
\includegraphics[width = 0.9\textwidth]{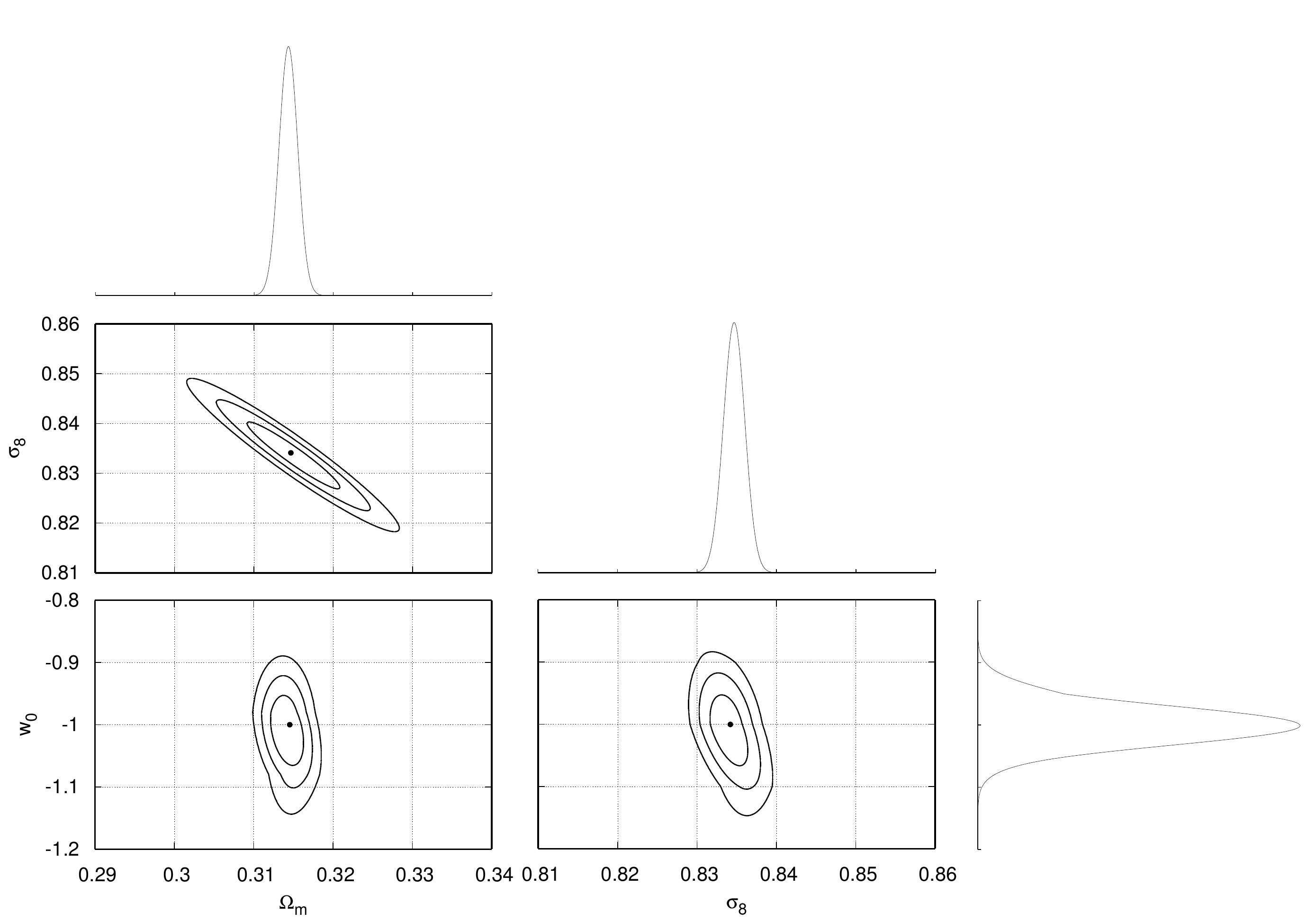}
\caption{Confidence regions for a survey with $A = 15000\,$deg$^2$) and two redshift bins. Confidence regions shown in the lower triangle are again $1\sigma$, $2\sigma$ and $3\sigma$. The plots on the diagonal show one dimensional cuts through the likelihood, while plots in the lower triangle show two dimensional cuts and the respective confidence regions. The fiducial cosmology is marked with a dot.}
\label{Fig:9}
\end{center}
\end{figure*}
We show in \autoref{Fig:8} the resulting likelihood without using redshift information in the $\Omega_\text{m}-\sigma_8$ plane. In this case we only fit $\Omega_\text{m}$ and $\sigma_8$ and keep $w_0$ fixed, because its variation around the fiducial value only marginally affects the results. The fiducial cosmology is marked with a point.
It can be seen that the expected degeneracy between $\Omega_\text{m}$ and $\sigma_8$ exists. Note that the likelihood is really non-Gaussian even though the contours look like ellipses, with contours corresponding to $68.3\%$, $95.4\%$ and $99.7\%$ confidence. The error contours are nearly orthogonal to those obtained from the CMB (\citealt{collaboration2013Pla201resXVICospar}) and thus a joint analysis can yield very tight constraints on both parameters. The relative uncertainties of both parameters are $\sim 10^{-3}$. Note that, assuming a source background density $n_g = 30\, \text{deg}^{-2}$ the relative uncertainty will pick up a factor of $1.2$ as mentioned before. Especially for $\sigma_8$ this result is competitive, for example \citet{2010MNRAS.409..737W} showed that a relative uncertainty of $\sim 7\times 10^{-2}$ can be achieved by using Baryon Acoustic Oscillation measurements.

So far we neglected any redshift information regarding the detections discarding all the information about the time evolution of the universe. Here, we introduce only two redshift bins for simplicity in order to trace the evolution of structure formation. By introducing redshift boundaries when integrating over Eq. (\ref{eq:12}). We would like to stress that here we assume the minimum amount of information when including redshift in the analysis, i.e. we do not assume to have the redshift of each individual detection, but we simply perform the peak counts on two shear maps, one for the higher and one for the lower redshift bin. This is because a redshift estimate can only be associated to those detections which have been identified to be galaxy clusters (\citealt{2011MNRAS.413.1145B}, \citealt{2012MNRAS.422..553B}). The estimate could be given by taking advantage of optical filters or other external data. Thus the procedure described in Sect. \ref{sec:5} has to be applied to each of the two shear maps separately. i.e. by fixing the total survey area one can calculate the optimal number of sub-fields for each shear map.

In order to get comparable statistics in the redshift bins, the expected number of objects should also be comparable. For our purpose we use two redshift bins such that the expected number of peaks above the threshold $t_{sn}$ is roughly equal, resulting into redshift bins with $0<z<0.35$ (named 'low') and $0.35<z$ (named 'high').

We fit $\Omega_\text{m}$, $\sigma_8$ and $w_0$ simultaneously while setting $w_\text{a}=0$. In \autoref{Fig:9} we show the resulting likelihoods. On the diagonal the maximized likelihoods can be seen, i.e. the one dimensional sub-spaces with the other parameters set to their best fit value. The lower triangle shows the maximized constraints on all three parameter permutations. One can see that the constraining power in the $\Omega_\text{m}-\sigma_8$ plane is very similar to the one obtained from \autoref{Fig:8}. However, $w_0$ can now be constrained allowing an uncertainty below $0.1$ at $1\sigma$ and thus a relative uncertainty $<10\%$. We also note that the degeneracy between $w_0$ and each of the other two parameters is very small. 

\section{Conclusions}\label{sec:7}
The results presented show that applying order statistics to the
distribution of cosmic shear peaks is a powerful tool to constrain
cosmological parameters associated with structure growth and the
geometrical evolution of the universe. We evaluated the extreme value
and order statistics for both over-densities and under-densities of
weak lensing shear maps, we optimized the criterion to split a wide
field survey into sub-fields in order to optimally sample the
distribution of extreme values, and we forecasted the constraints on
$\Omega_\text{m}$, $\sigma_8$ and $w_0$ achievable with wide fields
surveys. In the process we also improved the analytic recipe provided
in \citet{maturi2009anaapptonumcouweapeadet} and
\citet{2011MNRAS.416.2527M} by accounting for the impact of LSS on the
larges $S/N$ peaks which cannot be neglected even though these are caused by
highly non-linear structures. A good agreement was found when comparing
our model with a ray-tracing $N$-body numerical simulation.

This has been done with mock data under the assumption that (1) the
contribution of noise and LSS in weak lensing shear maps produced with
relatively broad kernels are well represented by a Gaussian random
field and (2) by ignoring blending between the peaks mainly caused by
noise and LSS and those caused by non-linear structures. For a Euclid
like survey with and without including photometric redshifts of the galaxies in our analysis we obtained the following:
\begin{enumerate}
\item When modelling the statistics of weak lensing peak maps the
  embedding of the contribution from clusters into the LSS has to be
  taken into account. While this might appear as counter-intuitive at first glance, but both LSS and noise fluctuations (even if
  occuring at low $S/N$ ratios) are very important also in the high
  signal-to-noise regime. Their impact is not negligible, especially
  when constraining the extreme value statistics, and cannot be
  ignored. The impact of the LSS and noise contributions to large $S/N$
  ratio peaks is of order 15\%. For this reason great attention has
  to be given to a detailed description of the data noise which, even
  if of low amplitude, may result in large biases if its statistical
  properties are not well understood.

\item Using extreme value statistics we give an analytic prediction
  for the largest weak lensing signals expected in wide field
  surveys. These values can also be used to verify if the existence of
  extremely massive clusters such as `El gordo' are falsifying
  $\Lambda$CDM or not. For example, this cluster, which has been
  widely claimed to be 'troublesome' for $\Lambda$CDM, is not in
  contrast expectations from $\Lambda$CDM (\citealt{2012MNRAS4201754W}). 

\item The extreme value statistics applied to peak counts has the
  advantage that the resulting likelihood from which cosmological
  parameters are extracted does not need any assumption of the shape of the distribution as it is given by the model itself. Furthermore the identification of the highest peaks is a
  straightforward task. 

\item We evaluated the constraints of the cosmological parameters
  achievable with the extreme value statistics applied to weak lensing
  maps. Booth $\Omega_\text{m}$ and $\sigma_8$ can be well
  constrained below the percent level, additionally the parameter degeneracy is nearly orthogonal to that obtained with CMB measurements.

\item The dark energy equation of state can also be constrained by
  splitting the ellipticity catalog in two redshift bins so that the
  absolute error of $w_0$ is $\Delta (w_0)\lesssim 0.1$. This result is
  competitive with other cosmological probes, e.g. the constraints
  obtained by the \citet{collaboration2013Pla201resXVICospar}.
\end{enumerate}

Investigating the highest shear peaks in wide weak lensing surveys, the
presented method directly reflects the statistics of the most massive
objects, i.e. galaxy clusters, avoiding any reference to their mass
and the danger of using potentially biased mass proxies and scaling
relations. Even though few assumptions have been used, this method has
the potential to provide strong constraints on the underlying
cosmological model.

\section*{Acknowledgments}
We would like to thank Bj\"orn Sch\"afer, Robert Lilow, Elena Kozlikin, Celia Viermann and Jonas Frings for helpful discussions. This work was supported in part by the Transregio-Sonderforschungsbereich TR 33
of
the Deutsche Forschungsgemeinschaft.

\bibliography{Paper}

\appendix
\section{Derivation of the GEV parameters}\label{app:1}
Equating coefficients of the Taylor expansion around the peaks of both distributions one finds via 
\begin{equation}
P_0(x_0) = G_{\gamma,\alpha,\beta}(x_0)
\end{equation}
the equation
\begin{equation}\label{eq:a1}
-n(x_0) A = -\left[1+\gamma\left(\frac{x_0-\alpha}{\beta}\right)\right]^{-1/\gamma}.
\end{equation}
The second term
\begin{equation}
P'_0(x_0) = G_{\gamma,\alpha,\beta}'(x_0)
\end{equation}
yields immediately
\begin{equation}\label{eq:a2}
-n'(x_0)A=\frac{1}{\beta}\left[1+\gamma\left(\frac{x_0-\alpha}{\beta}\right)\right]^{-1/\gamma-1}.
\end{equation}
Due to the maximum constraint the third term of the expansions has to vanish:
\begin{equation}
G_{\gamma,\alpha,\beta}''(x_0)\overset{!}{=}0 = P_0''(x_0)
\end{equation}
yielding
\begin{equation}\label{eq:a3}
1+\gamma = \left[1+\gamma\left(\frac{x_0-\alpha}{\beta}\right)\right]^{-1/\gamma}
\end{equation}
and Eq. (\ref{eq:19}). Plugging (\ref{eq:a3}) into (\ref{eq:a1}) yields 
\begin{equation}
\gamma = n(x_0)A-1.
\end{equation}
After inserting (\ref{eq:a3}) into (\ref{eq:a2}) one can conclude
\begin{equation}
\beta = -\frac{(1+\gamma)^{\gamma+1}}{n'(x_0)A}.
\end{equation}
Using again (\ref{eq:a2}) and inserting $\beta$ from the latter equation one ends up with
\begin{equation}
\alpha = x_0 -\frac{\beta}{\gamma}\left[(1+\gamma)^{-\gamma}-1\right].
\end{equation}

\bsp

\label{lastpage}

\end{document}